\documentclass[12pt]{article}

\usepackage{putex}
\usepackage{graphicx}
\usepackage{caption}
\usepackage{amsmath}
\usepackage{amssymb}
\usepackage{array}
\usepackage{bm}
\usepackage{multirow}
\usepackage{mathtools}
\usepackage{comment}
\usepackage{subcaption}
\usepackage{enumerate}
\usepackage{cite}
\usepackage{youngtab}
\usepackage{tensor}
\usepackage{slashed}
\usepackage[aligntableaux=center]{ytableau}
\usepackage[utf8]{inputenc}
\usepackage{rotating}
\usepackage{bigfoot}
\usepackage[
      colorlinks=true,
      linkcolor=blue,
      urlcolor=blue,
      filecolor=black,
      citecolor=red,
      linktocpage=true
      ]{hyperref}
\usepackage{dsfont}

\newcommand {\be} {\begin {equation}}
\newcommand {\ee} {\end {equation}}
\newcommand {\nn} {\nonumber}

\newcommand {\bes} {\begin {equation*}}
\newcommand {\ees} {\end {equation*}}

\newcommand{\es}[2]{%
  \begin{equation}
    \begin{aligned}
      #2
    \end{aligned}
    \phantomsection\label{#1}%
  \end{equation}%
}

\newcommand{\Z}{\mathbb{Z}}
\newcommand{\N}{\mathcal{N}}
\newcommand{\R}{\mathbb{R}}

\newcommand{\cA}{{\mathcal A}}
\newcommand{\cB}{{\mathcal B}}
\newcommand{\cC}{{\mathcal C}}
\newcommand{\cD}{{\mathcal D}}

\newcommand{\cG}{{\mathcal G}}

\newcommand{\cL}{{\mathcal L}}

\newcommand{\cO}{{\mathcal O}}

\newcommand{\cM}{{\mathcal M}}

\renewcommand{\L}{{\mathcal L}}

\newcommand{\rmd}{\mathrm{d}}

\newcommand{\bea}{\begin{equation}\begin{aligned}}
\newcommand{\eea}[1]{\label{#1}\end{aligned}\end{equation}}

\newcommand{\beq}{\begin{equation}}
\newcommand{\eeq}{\end{equation}}

\def\ie{\begin{equation}\begin{aligned}}
\def\fe{\end{aligned}\end{equation}}

\newcommand{\B}{{\mathcal{B}}}

\numberwithin{equation}{section}

\def\<{\langle}
\def\>{\rangle}

\usepackage{tikz}
\usetikzlibrary{decorations.pathmorphing, decorations.markings}
\usepackage{subcaption}

\begin{document}

\preprint{}

\institution{imperial}{Abdus Salam Centre for Theoretical Physics, Imperial College London, London SW7 2AZ, UK}
\institution{oxford}{Mathematical Institute, University of Oxford,
Woodstock Road, Oxford, OX2 6GG, UK }

\title{Extremal couplings and gluon scattering in M-theory}

\authors{Shai M.~Chester,\footnote{\texttt{s.chester@imperial.ac.uk}}\worksat{\imperial} Rishi Mouland,\footnote{\texttt{r.mouland@imperial.ac.uk}}\worksat{\imperial} Jesse van Muiden,\footnote{\texttt{j.van-muiden@imperial.ac.uk}}\worksat{\imperial} and Cl\'ement Virally\footnote{\texttt{clement.virally@maths.ox.ac.uk}}\worksat{\oxford}}

\abstract{
We consider M-theory on the backgrounds AdS$_4\times S^7/\mathbb{Z}_{N_f}$ and AdS$_7\times S^4/\mathbb{Z}_2$, which have fixed point locii AdS$_{d+1}\times S^3$ for $d=3,6$. These theories are holographically dual to certain CFTs in $d=3,6$ with eight supercharges. We compute the bulk cubic couplings between graviton KK modes and gluon KK modes living on the fixed points of these theories, which are generically extremal. We use these couplings to compute the graviton exchange term that appears in the strong coupling expansion of holographic correlators of gluon KK modes $\langle 22pp\rangle$ in these theories, and check that it matches the expected flat space limit. We express the answer in terms of a new reduced correlator solution to the superconformal Ward identities, which we derive for all CFTs with eight supercharges in $3\leq d\leq6$.
}
\date{}

\maketitle

\tableofcontents
\newpage

\section{Introduction}\label{sec:intro}

The AdS/CFT correspondence relates the cubic couplings $\beta_{ijk}$ of bulk fields to the OPE coefficient $\lambda_{ijk}$ of operators $\cO$ in the dual conformal field theory (CFT). If the scaling dimensions of these operators are related as $\Delta_i=\Delta_j+\Delta_k+2a$ for $a=0,1,2\dots$, then $\lambda_{ijk}$ seemingly diverges even if $\beta_{ijk}$ is finite, in which case we call $\beta_{ijk}$ an extremal coupling (for $a=0$) or super-extremal (for $a>0$) coupling. This divergence is related to mixing between $\cO_i$ and the composite operator $:\!\cO_j\square^{a}\cO_k\!:$ \cite{Castro:2024cmf}, such that CFT data of unmixed operators are all finite as expected.

Some of the simplest models where extremal couplings occur are theories with half-maximal supersymmetry where the bulk geometry has a singularity with a fixed point locus given by super-Yang-Mills (SYM) on AdS$_{d+1}\times S^3$ with gauge group $G_F$, which exist for $3\leq d\leq6$, see e.g. \cite{Alday:2021odx} for a review. The dual CFTs have an R-symmetry group that includes an $SU(2)_R$ factor,\footnote{In $d=5,6$ this is the entire group, while in 4d there is an extra $U(1)$ factor, and in 3d an extra $SU(2)$ factor, neither of which are relevant to the discussion in this paper.} and a flavor symmetry group $SU(2)_L\times G_F$. The $(d+4)$-dimensional SYM fields can be KK reduced to give an infinite tower of gluon modes dual to scalar single trace superprimaries with $\Delta=\epsilon p$ for $\epsilon\equiv \frac{d-2}{2}$ and $p=2,3,\dots$. These modes transform in the adjoint of $G_F$ and the $(\frac p2-1,\frac p2)$ of $SU(2)_L\times SU(2)_R$.\footnote{We denote irreps of $SU(2)_L\times SU(2)_R$ by their isospin.} The graviton fields in the full ambient spacetime of dimension $D$, which is either $11$ for M-theory or $10$ for string theory, can also be compactified to give several towers of graviton modes. The tower we study in this paper transform in the $(\frac r2-1,\frac r2-1)$, have $\Delta=\epsilon k_r$ for $k_r=r,r+2,r+4,\dots$, and are singlets under $G_F$. For instance, $k_2=2$ corresponds to the $(d+1)$-dimensional graviton itself that is dual to the protected stress tensor multiplet superprimary, while $k_2>2$ are dual to long multiplets. The cubic couplings $\beta_{pqk_r}$ between two gluon modes and a graviton mode is nonzero for $r=|p-q|+2,\dots,p+q-2$, and is generically (super-)extremal, which is related to mixing between the single trace graviton modes and double traces of gluon modes.

The cubic coupling can then be used to compute the contribution to gluon scattering $\langle 22pp\rangle$ from a graviton exchange term $M_R$, which gets contributions from the exchange of graviton modes for all $k_2=2,4,\dots$ in the direct channel and $k_p=p,p+2,p+4,\dots$ in the cross channel. The flat space limit \cite{Penedones:2010ue} of this AdS graviton exchange term is given by a $(d+4)$-dimensional flat space graviton exchange term \cite{Chester:2023qwo}:
\es{flatGrav}{
\cA_R(s,t)\sim\int\frac{d^{D-d-4}p_\perp}{(2\pi)^{D-d-4}}\left[\delta^{AB}\delta^{CD}\frac{1}{p_\perp^2-s}+\delta^{AC}\delta^{BD}\frac{1}{p_\perp^2-u}+\delta^{AD}\delta^{BC}\frac{1}{p_\perp^2-t}\right]\,,\\
}
where $s,t,u$ are Mandelstam variables, $A,B,C,D$ are adjoint indices for $G_F$, and we integrate over the transverse momentum between the $(d+4)$-dimensional SYM and the ambient spacetime $D=10,11$.

The cubic couplings and the resulting graviton exchange term were computed for the first time in the $d=4$ case in \cite{Chester:2025wti}. The bulk dual in this case is $N$ D3 branes in type IIB string theory probing various F-theory singularities \cite{Sen:1996vd,Dasgupta:1996ij}, with a low energy effective theory given by supergravity on AdS$_5\times S^5$ with certain singularities. The large $N$ expansion of $\langle22pp\rangle$ is fixed by the analytic bootstrap (i.e. crossing, analyticity, and the flat space limit) to take the form
\es{M4d}{
M_{4d}=&\frac{M_{F^2}}{N}+\frac{1}{N^2}\left[M_R+M_{F^2|F^2}(s,t)+\sum_i b^i M_{F^4,i }\right] +O\big(N^{-2}{\log N}\big)\,,\\
}
where $M_{F^2}$ is the tree level gluon exchange given in \cite{Alday:2021odx}, $M_{F^2|F^2}$ is the 1-loop gluon exchange term given in \cite{Alday:2021ajh}, and $M_{F^4,i}$ are contact terms with unknown coefficients $b_i$ due to higher derivative corrections $F^4$.\footnote{For $p=2$, the $b_i$ were fixed using supersymmetric localization in \cite{Chester:2025wti,Behan:2023fqq} for the unique F-theory compactification where the complexified string coupling $\tau$ can take any value, and the dual CFT thus has a Lagrangian.} In this case $\cA_R$ in \eqref{flatGrav} is logarithmically divergent, just like the flat space 1-loop gluon exchange term. After computing $\beta_{pqk_r}$, \cite{Chester:2025wti} found that the resulting $M_R$ matched the expected $\cA_R$ in the flat space limit with the correct coefficient. The graviton exchange term was then used to unmix the graviton modes and double traces of gluon modes at order $1/N$, which was particularly non-trivial due to the contribution of the $M_{F^2}$ term.

In this paper, we generalize this calculation to M-theory duals for $d=3,6$. The bulk dual in the first case is $N$ M2 branes probing a $\mathbb{C}^2/\mathbb{Z}_{N_f}$ orbifold singularity \cite{Acharya:1998pm,Acharya:2004qe,Anderson:2006pb}, with a low energy effective theory given by supergravity on AdS$_4\times S^7/\mathbb{Z}_{N_f}$ with fixed point locus AdS$_4\times S^3$ \cite{Ferrara:1998vf,Gomis:1998xj,Entin:1998ub,Pelc:1999ms}. The dual CFT is a 3d $U(N)$ gauge theory coupled to an adjoint hypermultiplet and ${N_f}$ fundamental hypermultiplets with $G_F=SU({N_f})$ \cite{Benini:2009qs,Bashkirov:2010kz}.\footnote{The theory also has an extra $U(1)$ flavor symmetry corresponding to monopole operators, but the gluon modes we consider are invariant under this symmetry, and so it is not relevant to our study.} The large $N$ expansion of $\langle22pp\rangle$ is fixed by analytic bootstrap to be\footnote{Analytic bootstrap constraints also allow for a $F^4$ contact term that scales like $N^{-7/6}$, but it was shown in \cite{Chester:2023qwo} for $\langle 2222\rangle$ that localization constraints fix it to zero since no such term appears in the standard matrix model expansion that appears in those constraints, which trivially extends to $\langle 22pp\rangle$. These localization constraints were also used to fix the $N^{-3/2}\log N$ terms for $\langle2222\rangle$, but they are insufficient to fix the general $\langle22pp\rangle$ case.}
\es{largeN3d}{
M_{3d}&=\frac{M_{F^2}}{c_J}+\frac{M_{F^2|F^2}}{c_J^2}+\frac{M_{R}}{c_T}+\frac{M_{F^2|F^2|F^2}}{c_J^3}+\frac{1}{N^{3\over 2}}\sum_{i}\kappa_i{M_{D^2F^4,i}}+O(N^{-3/2}\log N)\,,
}
where here $c_J\sim \sqrt{N}$ and $c_T\sim N^{3/2}$. The tree gluon exchange $M_{F^2}$ was computed in \cite{Alday:2021odx}, while the 1-loop gluon exchange $M_{F^2|F^2}$ is still unknown. The graviton exchange term $M_R$ now appears at the same order in $1/N$ as the 2-loop gluon exchange $M_{F^2|F^2|F^2}$ and the $M_{D^2F^4,i}$ contact terms. We compute $\beta_{pqk_r}$ for this theory and use it to compute $M_R$, which matches the expected $\cA_{R}$ in the flat space limit. Note that $\cA_R$ is logarithmically divergent just as in the 4d case, as can be seen from the four-dimensional transverse integral in \eqref{flatGrav}, except now it is the 2-loop gluon exchange that is also logarithmically divergent, while the 1-loop gluon exchange is not. 

The bulk dual in the $6d$ case is $N$ M5 branes probing an M9 end of the world brane \cite{Ganor:1996mu,Seiberg:1996vs}, with a low energy effective theory given by supergravity on AdS$_7\times S^4/\mathbb{Z}_2$ with fixed point locus AdS$_7\times S^3$. The dual CFT is a 6d $(1,0)$ CFT called E-string theory with $G_F=E_8$. The large $N$ expansion of $\langle22pp\rangle$ is fixed by analytic bootstrap to be
\es{largeN6d}{
M_{6d}&=\frac{M_{F^2}}{c_J}+\frac{M_{R}}{c_T}+O(N^{-10/3} )\,,
}
where here $c_J\sim N^2$ and $c_T\sim N^3$. The tree gluon exchange $M_{F^2}$ was computed in \cite{Alday:2021odx}. The graviton exchange term $M_R$ is now the first correction to $M_{F^2}$, and there are no contact term ambiguities at the same order in $1/N$. We compute $\beta_{pqk_r}$ for this theory and use it to compute $M_R$, which matches the expected $\cA_{R}$ in the flat space limit. Note that $\cA_R$ is now finite, as can be seen from the one-dimensional transverse integral in \eqref{flatGrav}, which is related to the lack of contact term ambiguities in this case.

A key technical ingredient to our computation is a novel way of solving the Ward identities that encode the constraints of superconformal symmetry on $\langle22pp\rangle$ in general $3\leq d\leq6$ \cite{Dolan:2004mu}. In even $d$, it was shown in the original paper that these constraints can be formally solved in position space by a writing the correlator in terms of a certain differential operator acting on an unconstrained function called the reduced correlator. This differential operator is well defined in even $d$, where the reduced correlator has its own expansion in superconformal blocks \cite{Dolan:2004mu,Chang:2017xmr,Beem:2014zpa}. In odd $d$, however, the differential operator involves fractional derivatives. In this paper, following \cite{Virally:2025nnl}, we show how the differential operator can be made well defined in any $d$ by going to Mellin space. The resulting reduced correlator then has an expansion in the Mellin space analogue of superblocks,\footnote{For every multiplet except for the flavor multiplet, as we discuss in more detail in the main text.} which correspond to Witten exchange diagrams. This expansion of the reduced correlator was essential in efficiently resumming the infinite exchange diagrams that contribute to the graviton exchange term $M_R$, and we expect should be useful in other contexts.

The rest of this paper is organized as follows. In Section \ref{sec:sugra} we consider the effective bulk theory of 11d supergravity coupled to $(d+4)$ dimensional SYM, and use this compute the bulk couplings $\beta_{pqk_r}$ for $d=3,6$. In Section \ref{sec:corr}, we discuss constraints from superconformal symmetry on $\langle 22pp\rangle$, including our novel reduced correlator in Mellin space for general $3\leq d \leq 6$. In Section \ref{sec:scattering} we discuss gluon scattering $\langle 22pp\rangle$ for $d=3,6$, including the graviton exchange term and mixing between the graviton modes and double traces for $d=6$. We conclude in Section \ref{sec:conclusion} with a review of our results and a discussion of future directions. Technical details of the calculations are given in the various Appendices. We also include an attached \texttt{Mathematica} notebook with certain lengthy equations.

\section{Extremal couplings in M-theory on AdS}\label{sec:sugra}
To compute the bulk couplings in M-theory, we start by considering the eleven-dimensional supergravity action\footnote{Our Hodge dual convention is such that for a $p$-form $\alpha$ we have $(\star \alpha)_{a_1\dots a_{11-p}}=\frac{1}{p!}\epsilon_{b_1\dots b_p a_1\dots a_{11-p}\alpha^{b_1\dots b_p}}$ in terms of volume form $\epsilon_{1\dots 11}=\sqrt{-g}$. Note that $\star^2 \alpha_p=-\alpha_p$.}
\begin{equation}
	S_{\text{11}} = \frac{1}{2\kappa^2} \int_{M_{11}} \left( R \star 1 - \frac{1}{2} G_4 \wedge \star \,G_4 + \frac{1}{6} C_3 \wedge G_4 \wedge G_4 \right) \,,\quad \text{with}\quad G_4 = \rmd C_3\,,
\label{eq: 11d SUGRA}
\end{equation}
where $R$ is the eleven-dimensional Ricci scalar, $C_3$ is the 3-form field, and the gravitational coupling is related to the eleven-dimensional Planck length in the standard way $2\kappa^2 = (2\pi)^{8} \ell_p^9$. The particular geometries of interest have the following metrics
\begin{equation}
	\rmd s_{11}^2 = L^2 \left( \rmd s_{\text{AdS}_{d+1}}^2 + \epsilon^{-2} \,\rmd s_{S^{n}/\Z_{N_f}} \right)\,,
\label{eq: metric}
\end{equation}
where $d+1+n=11$ and we are interested in the following two cases: $d=3$ with ${N_f}$ generic, and $d=6$ with ${N_f}=2$.\footnote{In the $d=6$ case $N_f=2$ is not to be interpreted as a number of flavor multiplets in the field theory.} The parameter $\epsilon$ fixes the difference in length scales between the compact and non-compact spaces and was defined already in the introduction to equal
\begin{equation}
	\epsilon = \frac{d-2}{2}\,.
\end{equation}
Note that $ds_{S^n/\Z_{N_f}}$ is just the unit round metric on $S^n$ subject to an orbifold we describe below. We set $L=1$ in all that follows. The background form field fluxes are
\begin{align}
  G_n = \frac{d}{\epsilon^n} \text{vol}_{S^n/\Z_{N_f}}\quad \longleftrightarrow \quad G_{d+1} = d(-1)^d  \text{vol}_{\text{AdS}_{d+1}}\,,
  \label{eq: fluxes}
\end{align}
where we write $G_7 = \rmd C_6 = \star G_4- \frac{1}{2}C_3\wedge G_4$.  

Let us then describe the $\Z_{N_f}$ orbifold. We can view $S^n$ as an $S^3\times S^{n-4}$ fibration over an interval, where the $S^3$ shrinks at one end, and the $S^{n-4}$ at the other. Then $\Z_{N_f}$ acts freely around the $S^{n-4}$ fibre, and so in particular has an $S^3$ fixed point locus. Concretely, we can take the metric
\begin{equation}
	\rmd s_{S^n/\Z_{N_f}}^2 = \rmd \theta^2 + \sin^2 \theta \,\rmd s_{S^{3}}^2 + \cos^2 \theta \, \rmd s_{S^{n-4}/\Z_{N_f}}^2\,,
\end{equation}
where the metric on $S^{n-4}/\Z_{N_f}$ is locally just the unit round metric on $S^{n-4}$. The fixed point locus is then at $\theta=\pi/2$. Note that $S^0=\{\pm 1\}$ and so for the $d=6$ case we can only take a $\Z_2$ orbifold as mentioned above.

The isometries of the internal space are thus broken as 
\begin{equation}
	\text{SO}(n+1) \rightarrow \text{SU}(2)_L \times \text{SU}(2)_R \times H\,,
\label{eq: symmetric breaking}
\end{equation}
where $SO(4) \cong SU(2)_L\times SU(2)_R$ rotate the $S^3$ while $H\subset SO(n-3)$ is the subgroup of rotations of $S^{n-4}$ that is preserved by the orbifold. In particular for our $d=3$ case the $\Z_k$ acts along the Hopf fibre in $S^{n-4}=S^3$ and thus $H=U(1)\times SU(2)$ where the $SU(2)$ factor is the additional R-symmetry mentioned above, while the $U(1)$ is a flavor symmetry. Meanwhile for the $d=6$ case $H$ is trivial.

The value of the gravitational coupling $\kappa$ in terms of the dual field theory rank $N$ is given in Table \ref{Tab: relevant constants}, where recall we are setting the AdS radius $L=1$. This is determined by the flux quantisation condition
\begin{align}
  \frac{1}{(2\pi l_p)^{n-1}}\int_{S^n/\Z_{N_f}} G_n = N\,.
\end{align}
These backgrounds harbour Yang-Mills degrees of freedom localised at the AdS$_{d+1}\times S^3$ fixed point locus. The effective action describing these modes is
\begin{align}
  S_\text{brane} = \frac{1}{g_\text{YM}^2}\, \text{Tr} \int_{\text{AdS}_{d+1}\times S^3}\Big(-F\wedge \star_{d+4} F + C_{d}\wedge F \wedge F\Big)\,,
  \label{eq: brane action}
\end{align}
where $\star_{d+4}$ denotes the Hodge dual with respect to the pullback of the metric (\ref{eq: metric}) to AdS$_{d+1}\times S^3$, using the same conventions as above.

In the AdS$_4$ case, $g_{\text{YM}}^2$ (i.e. the inverse brane tension) was fixed in \cite{Chester:2023qwo}, and is given (in terms of $N$) in Table \ref{Tab: relevant constants}. Meanwhile in the AdS$_7$ case we fix $g_{\text{YM}}^2$ below using the relation between $g_{\text{YM}}^2$ and the flavor central charge $c_J$ of the dual field theory, along with the independent determination of $c_J$ as a function of $N$ in the field theory; the result is also given in Table \ref{Tab: relevant constants}.

\begin{table}
\centering
\renewcommand{\arraystretch}{1.5}
\begin{tabular}{c|ccccccccc}
	 Background & $\kappa^2$ & $g_\text{YM}^2$ & $c_T$ & $c_J$\\\hline
AdS$_4\times S^7/\Z_{N_f}$   & $\frac{256\sqrt{2}\pi^5}{ ({N_f}N)^{3/2}}$  & $\frac{32\sqrt{2} \pi^3}{ ({N_f}N)^{1/2}}$ &$\frac{32\sqrt{2{N_f}}}{\pi}N^{3/2}+\cO(N^{1/2})$ & $\frac{8\sqrt{2 {N_f}}}{\pi}N^{1/2}+\cO(N^0)$\\
AdS$_7\times S^4/\Z_2$    & $\frac{\pi^5}{4(2N)^3}$ & $\frac{2\pi^5}{(2N)^2}$ & $1344 N^3 + \cO(N^2)$	& $ 60N^2 + \cO(N)$
\end{tabular}
\caption{Summary of constants relevant for the AdS$_{d+1} \times S^n/\mathbb{Z}_{N_f}$ backgrounds of interest.}\label{Tab: relevant constants}
\end{table}
These degrees of freedom are understood in the $d=3$ case as M2 branes sourcing a KK-monopole, while in the $d=6$ case they live on an end-of-the-world brane. We note that the full Wess-Zumino terms in these effective actions take a more complicated form than what we present in \eqref{eq: brane action} \cite{Bergshoeff:1998ef,Bergshoeff:1998re,Eyras:1998rf}, nevertheless this one term is the only relevant piece that will play a role in our analysis. 

%
%

%
\subsection{Central charges}
As a consistency check, let us verify that our bulk setup reproduces the known values of the dual field theory central charges. These central charges are defined through the two-point functions of their associated currents
\begin{equation}
\begin{aligned}
\label{charges}
	\left< T_{ab}(z) T_{cd}(0)\right> =&\, \frac{\Gamma(d/2)^2}{4\pi^{d}} \frac{c_T}{z^{2d}} \left( I_{a(c}I_{d)b} - \text{trace} \right)\,,\\
	\left< J^A_{a}(z) J^B_{b}(0)\right> =&\, \frac{\Gamma(d/2)^2}{4\pi^{d}}\frac{c_J}{z^{2(d-1)}}  \delta^{AB} I_{ab}\,,	
\end{aligned}
\end{equation}
where $I_{ab} = \delta_{ab} - 2 \frac{z_a z_b}{z^2}$. These central charges are related straightforwardly in the bulk to the gravitational and Yang-Mills couplings by \cite{Liu:1998bu,Freedman:1998tz}\footnote{In these references the central charges are related to the effective $(d+1)$-dimensional couplings in the bulk, which we here directly rewrote into their higher-dimensional counterparts coming from \eqref{eq: 11d SUGRA} and \eqref{eq: brane action}.}
\begin{equation}
\begin{aligned}
  c_T =	\frac{4\pi^{d/2}\Gamma(d+2)}{(d-1)\Gamma(d/2)^3} \frac{\text{Vol}(S^n)}{\epsilon^n N_f} \frac{1}{\kappa^2}
  \,,\quad  c_J = \frac{2(d-2)\pi^{d/2}\Gamma(d)}{\Gamma(d/2)^3} \frac{\text{Vol}(S^3)}{\epsilon^3}\frac{1}{g_\text{YM}^2}
  \,.
\end{aligned}
\label{eq: cT and cJ}
\end{equation}
For $c_T$, we can then in both the AdS$_4$ and AdS$_7$ case plug in the value of $\kappa^2$ in terms of $N$, as given in Table \ref{Tab: relevant constants}, to determine $c_T$ at leading order at large $N$. The result is given in the same table, and in both cases, precisely matches field theory results in \cite{Chang:2017xmr} for AdS$_7 \times S^4/\Z_2$ and \cite{Chester:2020jay}\footnote{Note we use the conventions of \cite{Chang:2017xmr} for $c_T$, and so in particular versus \cite{Chester:2023qwo} we have $c_T^\text{ours} = \frac{3}{2}c_T^\text{theirs}$.} for AdS$_4 \times S^7/\Z_{N_f}$.

For $c_J$, in AdS$_4$ we know $g_\text{YM}^2$ (i.e. the inverse brane tension) independently in the bulk, and so can plug this into (\ref{eq: cT and cJ}) to determine the leading large $N$ value of $c_J$. This matches precisely the field theory result of \cite{Chester:2020jay}. Meanwhile in AdS$_7$, we use the known value of $c_J$ in the field theory \cite{Chang:2017xmr} to fix $g_\text{YM}^2$ using (\ref{eq: cT and cJ}), with the result given in Table \ref{Tab: relevant constants}.

\subsection{Cubic couplings}\label{subsec: cubic couplings}

We now want to compute the bulk couplings $\beta_{pqk_r}$ between the relevant Kaluza-Klein scalar fields in AdS$_{d+1}$. This calculation proceeds in a way analogous to that of \cite{Chester:2025wti}; as such, we sketch the basic steps here, relegating full details to Appendix \ref{app: cubic couplings}.

The first step is to identify the relevant fluctuations of the supergravity fields $(g,C_3)$. Let us first forget about the orbifold and just consider the AdS$_{d+1}\times S^n$ background. The task at hand then is to expand the action (\ref{eq: 11d SUGRA}) in fluctuations around the given background. Our interest is in fluctuations which give rise to scalar fields on AdS$_{d+1}$ upon dimensional reduction. The non-triviality comes in diagonalising the quadratic action of these fluctuations. This problem was studied in detail for AdS$_4\times S^7$ in \cite{Castellani:1984vv} and for AdS$_7\times S^4$ in \cite{vanNieuwenhuizen:1984iz}, which we review in some detail in our conventions in Section \ref{subsec: diag}. In both cases, scalar fields in AdS$_{d+1}$ are found to correspond to scalar spherical harmonics on $S^n$, which in turn are labelled by an integer $k$. For each $k$, one in fact finds a pair of such scalar modes, but only one gives rise after the orbifold to the superprimaries of interest, while the other gives rise to superdescendents. For each\footnote{For $k=0,1$ these modes are pure gauge.} $k\ge 2$, the fluctuation we need then takes the form
\begin{align}
  \delta g_{\mu\nu} 		&= a_1(k) g_{\mu\nu} s_k +  a_2(k)\left(\nabla_\mu \nabla_\nu  - \tfrac{1}{d+1}g_{\mu\nu}\square_\text{AdS}\right)s_k	\,,	\nn\\
  \delta g_{\alpha\beta}		&= a_3(k) g_{\alpha\beta} s_k	\,,	\nn\\
  \delta g_{\mu \alpha }		&= 0		\,,		\nn\\
  \delta G_{d+1}\big|_{\text{AdS}_{d+1}}			&= a_4(k)   s_k  	\text{vol}_{\text{AdS}_{d+1}} 	\,,
\label{eq: generic graviton fluctuation}
\end{align} 
where $\mu,\nu,\dots=1,\dots,d+1$ are indices in AdS$_{d+1}$ and $\alpha,\beta,\dots = 1,\dots, n$ are indices in $S^n$. In this expression, $s_k$ is a scalar spherical harmonic on $S^n$ transforming in the traceless symmetric rank $k$ representation of $SO(n+1)$. The derivation of the functions $a_i(k)$ is presented in Appendix \ref{subsec: diag}, and their explicit form can be found in \eqref{eq: a and b AdS4} and \eqref{eq: a and b AdS7}.


Plugging the fluctuation (\ref{eq: generic graviton fluctuation}) into the supergravity action (\ref{eq: 11d SUGRA}) we find at quadratic order
\begin{equation}
  S = b(k) \int d^{11} x \sqrt{-g}\left(- \frac{1}{2} \nabla_\mu s_k \nabla^\mu s_k - \frac{1}{2}\Delta_k (\Delta_k - d)s_k^2 \right)\,,
\label{eq: SUGRA kinetic term}
\end{equation}
where $b(k)$ is background-dependent and given in (\ref{eq: a and b AdS4}) and (\ref{eq: a and b AdS7}), and we find the expected dual scaling dimension
\begin{equation}
  \Delta_k = \epsilon k\,.
\end{equation}
The next step is to determine the fate of these excitations when we introduce the orbifold. This amounts to decomposing the rank $k$ traceless symmetric representation of $SO(n+1)$ under the breaking (\ref{eq: symmetric breaking}). Furthermore, since the gluon modes at the fixed point are uncharged under $H$ and in the adjoint of $G_F$, they can only admit cubic couplings with graviton modes that are neutral under $H$. So let's focus on these. The rank $k$ representation then contains a set of representations we denote $k_p$ where $p$ with $p=k+2,k,k-2,\dots$, ending at $p=2$ for $k$ even or $p=3$ for $k$ odd. This mode $k_p$ transforms with isospins $(\frac{p}{2}-1,\frac{p}{2}-1)$ under $SU(2)_L\times SU(2)_R$. The explicit expression for these harmonics is given in (\ref{eq: kp mode explicit}). 

The third step is to determine the relevant excitations of the AdS$_{d+1}\times S^3$ worldvolume Yang-Mills field $A$ corresponding to the scalar gluon modes to which the supergravity fluctuations described above couple. These modes transforms in the $(\frac{p}{2}-1,\frac{p}{2})$ and is given by a rank $p$ vector spherical harmonic on $S^3$,  described explicitly in (\ref{eq: Vp mode}).

We finally put all this together. Plugging into the action (\ref{eq: brane action}) the linear perturbations of the metric, form field and gauge field, we arrive at the following answer for the cubic couplings\footnote{Precisely how these coupling are normalised is described in (\ref{eq: S cubic}).}
\begin{equation}\label{Eq: cubic coupling general d n}
	\begin{aligned}
		&\beta_{pqk_r}^{} = 	\frac{\kappa \epsilon(d-2)(d+n-1)}{72 \,\Gamma(\frac{n-1}{2})}\sqrt{\frac{2\epsilon^n N_f}{\text{Vol}(S^n)}\frac{(p-1)(q-1)(r-1)}{k_r (k_r - 1)(d k_r + n - 1 )} \frac{\Gamma(\frac{k_r - r + n - 1}{2})\Gamma(\frac{k_r + r + n - 3}{2})}{\Gamma(\frac{k_r - r + 4}{2})\Gamma(\frac{k_r + r + 2}{2})} } \\
		& \times \frac{\Gamma(p - 1)\Gamma(q - 1)\Gamma(r - 1)(k_r + p - q)(k_r + q - p)(k_r + p + q - n + 1)(k_r + p + q - 2)}{\Gamma\left(\frac{p + q - r}{2}\right)\Gamma\left(\frac{p - q + r}{2}\right)\Gamma\left(\frac{q - p + r}{2}\right)\Gamma\left(\frac{p + q + r - 2}{2}\right)}\,.
	\end{aligned}
\end{equation}
Interestingly, we find that taking $(d,n,\epsilon) = (4,5,1)$ this result also reproduces the correct corresponding cubic coupling in AdS$_5 \times S^5/\Z_{N_f}$ computed in \cite{Chester:2025wti}.\footnote{This holds up to an overall constant related to the implementation of the orbifold number and length scales.}

To relate this cubic coupling directly to the boundary OPE coefficient one needs to introduce a factor coming from the vertex integral in the bulk such that \cite{Freedman:1998tz}
\begin{equation}
	\lambda_{ijk}=\beta_{ijk}\frac{\pi ^{d/2}\sqrt{\cC_{\Delta_i}\cC_{\Delta_j}\cC_{\Delta_k}}  \Gamma(\frac{\Delta_i+\Delta_j-\Delta_k}{2})  \Gamma(\frac{\Delta_k+\Delta_i-\Delta_j}{2})  \Gamma(\frac{\Delta_j+\Delta_k-\Delta_i}{2})
 \Gamma(\frac{\Delta_i+\Delta_j+\Delta_k-d}{2})
}{2 \Gamma(\Delta_i) \Gamma(\Delta_j) \Gamma(\Delta_k)}\,,
\end{equation}
where $\mathcal C_{\Delta} = \frac{\Gamma(\Delta)}{2\pi^{d/2}\Gamma(\Delta+1-d/2)}$. Note that this relation between bulk cubic couplings and boundary OPE coefficients only holds in the non-extremal case as this expression otherwise diverges. For future reference we will explicitly provide the expressions of the OPE coefficients in the two cases of relevance to us. Namely, for AdS$_4\times S^7/\Z_{N_f}$ we find that
%
\begin{equation}\label{lam3d}
\begin{aligned}
	\lambda_{pqk_r}^{d=3}&=\frac{ \Gamma (r-1) \Gamma \left(\frac{1}{4}
   (k_r+p-q+4)\right) \Gamma \left(\frac{1}{4} (k_r-p+q+4)\right)
   \Gamma \left(\frac{1}{4} (-k_r+p+q)\right) \Gamma \left(\frac{1}{2}
   (k_r+p+q+2)\right) }{ \Gamma \left(\frac{1}{2} (p+q-r)\right) \Gamma
   \left(\frac{1}{2} (p-q+r)\right) \Gamma \left(\frac{1}{2} (-p+q+r)\right)
   \Gamma \left(\frac{1}{2} (p+q+r-2)\right)\Gamma \left(\frac{1}{4}
   (k_r+p+q+4)\right)}\\
   &\times \frac{1}{4} \left( \frac{1}{2N_f N^3} \right)^{1/4} \sqrt{\frac{(r-1) (k_r-r+4) (k_r+r+2) \Gamma (p) \Gamma (q) }{\pi(k_r+2) \Gamma (k_r+1)}}+O(N^{-5/4})\,,	
\end{aligned}
\end{equation}
while for AdS$_7\times S^4/\Z_2$ instead we have that
\begin{equation}\label{lam6d}
\begin{aligned}
	\lambda_{pqk_r}^{d=6}&=\frac{\Gamma (p) \Gamma (q) \Gamma (r-1) \Gamma (k_r+p-q+1) \Gamma (k_r-p+q+1) \Gamma (-k_r+p+q) \Gamma (k_r+p+q-1) }{\Gamma (2 k_r) \Gamma (2 p) \Gamma(2 q) \Gamma \left(\frac{1}{2} (p+q-r)\right) \Gamma \left(\frac{1}{2}(p-q+r)\right) \Gamma \left(\frac{1}{2} (-p+q+r)\right) \Gamma \left(\frac{1}{2} (p+q+r-2)\right)}\\
   &\times \frac{1}{2 N^{3/2}} \sqrt{\frac{{(2 k_r-1) (2 p-1) (2 q-1) (r-1) \Gamma \left(\frac{1}{2} (k_r-r+3)\right) \Gamma \left(\frac{1}{2}(k_r+r+1)\right)}}{\pi{k_r (2 k_r+1) \Gamma \left(\frac{1}{2} (k_r-r+4)\right) \Gamma \left(\frac{1}{2} (k_r+r+2)\right)} }}+O(N^{-2})\,.	
\end{aligned}
\end{equation}
A consistency check can be found by taking $p=q$ and $k_2 = 2$, where it is expected that the OPE coefficients squared are related to the central charge $c_T$ with a factor quadratic in $p$ \cite{Osborn:1993cr}.\footnote{More generally, we have $\lambda^2_{\phi\phi T}\sim\Delta_\phi^2/c_T$, where $T$ is the stress tensor.} Using the relation between $N$ and $c_T$ for each theory as shown in Table \ref{Tab: relevant constants}, we indeed find that
\begin{equation}
	(\lambda_{pp (k_2 = 2)}^{d=3} )^2 = \frac{6p^2}{c_T}\,,\quad \text{and} \quad (\lambda_{pp (k_2 = 2)}^{d=6} )^2 = \frac{42p^2}{5 c_T}\,,
	\label{cTrel}
\end{equation}
where the precise coefficients in each case matches the general formula \cite{Chang:2017xmr}, which we will review in \eqref{identity}.\footnote{To compare to the 3d results in \cite{Chester:2023qwo} one has to take into account the different normalisation in $c_T$ as mentioned above, and the fact that the relevant flavor tensor structure normalisation in \cite{Chester:2023qwo} was taken to be $A^{ABCD} = \frac14 \delta^{AB} \delta^{CD}$, whereas we have chosen the unit normalisation as given in \eqref{proj}.}

\section{Half-BPS four-point functions}\label{sec:corr}

Our main application of the cubic couplings we just computed is to scattering of gluons and higher KK modes, which is dual to four point functions $\langle 22pp\rangle$. Here, $p$ denotes the scalar superprimary of a half-BPS operator with $\Delta=\epsilon p$ and $\epsilon=\frac{d-2}{2}$ in theories with eight supercharges in spacetimes $d=3,5,6$,\footnote{We do not discuss $d=4$, because that theory has a protected subsector described by a two dimensional chiral algebra \cite{Beem:2013sza}, which makes it very different from the generic case.} which all have an $SU(2)_R$ subsector of their R-symmetry. In this section, we will discuss general constraints on this correlator from the superconformal algebra. We start by reviewing the superblock expansion following \cite{Dolan:2001tt,Chang:2017xmr,Baume:2019aid,Bobev:2017jhk}. We will then discuss how the constraints of superconformal symmetry can be solved by a reduced correlator with its own block expansion, which was already known for $d=6$ \cite{Chang:2017xmr}, but is novel for $d=3,5$.

\subsection{Setup}
\label{setup}

We consider half-BPS multiplets $\mathcal{D}[p]$, whose superprimary is the Lorentz scalar $\phi_p^A(y,\bar y,x)$ with scaling dimension $\Delta=\epsilon p$, which transforms in the isospin $\frac p2$ irrep of $SU(2)_R$ with spinor polarization $y$. For instance, $p=2$ corresponds to the flavor multiplet. In the next section we will consider theories whose flavor symmetry is $SU(2)_L\times G_F$, so we also assume the superprimary transforms in the adjoint $\mathfrak{g}$ of $G_F$ with index $A=1,\dots,\dim(G_F)$, as well as the isospin $\frac p2-1$ irrep of $SU(2)_L$ with spinor polarization $\bar y$.\footnote{The analysis in this section only cares about the $R$-symmetry, so the other symmetries and their indices just go along for the ride.} The conformal and global symmetries restrict $\langle \phi_2\phi_2\phi_p\phi_p\rangle$ (denoted as $\langle22pp\rangle$) to be
\es{phiExp1}{
&\langle \phi_2^A(y_1,x_1) \phi_2^B(y_2,x_2) \phi_p^C(y_3,\bar y_3,x_3) \phi_p^D(y_4,\bar y_4,x_4) \rangle \\
&\hspace{20mm}= \frac{\langle y_1,y_2\rangle^2 \langle y_3,y_4\rangle^p \langle \bar y_3,\bar y_4\rangle^{p-2}}{x_{12}^{4\epsilon}x_{34}^{2p\epsilon}}\sum_{\bf r\in\mathfrak{g} \otimes \mathfrak{g}}G_{\bf r}(U,V;\alpha)P_{\bf r}^{ABCD}\,,\\
&\langle \phi_2^A(y_1,x_1) \phi_p^B(y_2,x_2) \phi_2^C(y_3,\bar y_3,x_3) \phi_p^D(y_4,\bar y_4,x_4) \rangle\\
&\hspace{20mm} = \frac{x_{24}^{(2-p)\epsilon}\langle y_1,y_2\rangle^{1+\frac p2} \langle y_3,y_4\rangle^{1+\frac p2} \langle \bar y_2,\bar y_4\rangle^{p-2} }{x_{12}^{(2+p)\epsilon}x_{34}^{(2+p)\epsilon}x_{13}^{(2-p)\epsilon}\langle y_1,y_3\rangle^{\frac p2-1}\langle y_2,y_4\rangle^{1-\frac p2}}\sum_{\bf r\in\mathfrak{g} \otimes \mathfrak{g}}\widetilde G_{\bf r}(U,V;\alpha)P_{\bf r}^{ABCD}\,,
}
where for later convenience we define both $G_{\bf r}$ in the $\langle22pp\rangle$ configuration and $\widetilde G_{\bf r}$ in the $\langle2p2p\rangle$ configuration. The projectors onto an irrep $\bf r$ of $G_F$ are normalized as\footnote{Our normalization for the identity projector differs from \cite{Chang:2017xmr} by factor of $\dim(G_F)$, which is chosen to make formulae for OPE coefficients with graviton modes simpler in what follows.}
\es{proj}{
P_{\bf r\neq1}^{ABBA}=\dim({\bf r})\,,\qquad P_{\bf 1}^{ABCD}=\delta^{AB}\delta^{CD}\,,
}
which are the only properties we will use in this work. The conformal cross ratios $U,V$ and the R-symmetry cross ratio $\alpha$ are defined as
 \es{crossRatios}{
 U\equiv \frac{\vec x_{12}^2 \vec x_{34}^2}{\vec x_{13}^2  \vec x_{24}^2} \equiv z \bar z \,,\qquad  V\equiv \frac{\vec x_{14}^2 \vec x_{23}^2}{\vec x_{13}^2  \vec x_{24}^2}\equiv (1-z)(1-\bar z) \,,\qquad \alpha=\frac{ \langle y_1,y_3\rangle  \langle y_2,y_4\rangle}{ \langle y_1,y_2\rangle  \langle y_3,y_4\rangle}\,.
 }
 The four point functions are then related by crossing $1\leftrightarrow2$ and $1\leftrightarrow3$ as
 \es{cross}{
 G^{ABCD}(U,V,\alpha)=G^{BACD}(U/V,1/V,1-\alpha)=U^{\epsilon\frac{p+2}{2}}\alpha^{\frac{p+2}{2}}\widetilde{G}^{ACBD}(1/U,V/U,1/\alpha)\,,
 }
 where for later convenience we wrote the $G_F$ indices explicitly, instead of in terms of projectors. Note that when $p=2$ we have $\widetilde{G}=G$, so there is an extra constraint on $G$.
 
The constraints from superconformal symmetry are encoded by the Ward identities \cite{Dolan:2001tt}
\es{ward}{
\left(z\partial_z-\epsilon\alpha\partial_\alpha\right){G}(U,V;\alpha)\big\vert_{\alpha=z^{-1}}=0\,,
}
where $z$ is defined as in \eqref{crossRatios}, and $G$ here stands for either $G_{\bf r}$ or $\widetilde G_{\bf r}$ in \eqref{phiExp1}, and a similar equation holds with $z\leftrightarrow \bar z$. The Ward identities can be solved by expanding the correlators as
\es{superblockExp}{
{G}_{\bf r}(U,V,\alpha)&=\sum_{\mathfrak{M}\in \mathcal{D}[2]\times  \mathcal{D}[2]}\lambda_{22(\mathfrak{M},{\bf r})}\lambda_{pp(\mathfrak{M},{\bf r})}\mathfrak{G}_\mathfrak{M}(U,V,\alpha)\,,\\
 \widetilde{G}_{\bf r}(U,V,\alpha)&=\sum_{\mathfrak{M}\in \mathcal{D}[2]\times  \mathcal{D}[p]}\lambda^2_{2p(\mathfrak{M},{\bf r})}\widetilde{\mathfrak{G}}_\mathfrak{M}(U,V,\alpha)\,,
} 
where $\lambda$ are OPE coefficients for each supermultiplet $\mathfrak{M}$ in $G_F$ irrep $\bf r$. The superblocks $\mathfrak{G}_\mathfrak{M}$ can be expanded in conformal blocks as
\es{superExp}{
{\mathfrak{G}}_\mathfrak{M}(U,V,\alpha)&=\sum_{J=0}^{2}\sum_{a=0}^4\sum_{b=-2}^2 {\mathcal{Y}}^2_J(\alpha)f^J_{\Delta+a,\ell+b}g^{0,0}_{\Delta+a,\ell+b}(U,V)\,,\\
\widetilde{\mathfrak{G}}_\mathfrak{M}(U,V,\alpha)&=\sum_{J=p/2-1}^{p/2+1}\sum_{a=0}^4\sum_{b=-2}^2 {\mathcal{Y}}^p_J(\alpha) f^J_{\Delta+a,\ell+b}g^{\epsilon(2-p),\epsilon(2-p)}_{\Delta+a,\ell+b}(U,V)\,,\\
}
where $g^{\Delta_{12},\Delta_{34}}_{\Delta,\ell}(U,V)$ are conformal blocks\footnote{We normalize our blocks such that $\lim_{U\to0,V\to1}g_{\Delta,\ell}\sim (1-V)^\ell U^{\frac{\Delta-\ell}{2}}$.} and $ \mathcal{Y}^p_J(\alpha)$ are degree $J$ polynomials\footnote{We normalize these polynomials as in \cite{Alday:2021odx} aside from an extra power of $\alpha^{\frac p2-1}$, such that both the $\alpha^{\frac p2-1}$ and $\alpha^{\frac p2+1}$ terms have unit coefficient.}
\es{Ys}{
\mathcal{Y}_J^p(\alpha)=\frac{\alpha^{\frac{p}{2}-1}}{(2 J)!}\Gamma \left(-\frac{p}{2}+J+2\right) \Gamma
   \left(\frac{p}{2}+J\right)
   P_{-\frac{p}{2}+J+1}^{(0,p-2)}(2 \alpha -1)\,,
}
which correspond to isospin $J$ irreps of $SU(2)_R$. The coefficients $f$ are given in \cite{Baume:2019aid}, and take a lengthy form.\footnote{Our conventions are related to their as $f^J_{\Delta,\ell,\text{us}}=(-1)^{\ell-\ell_0}\frac{\Gamma \left(\frac{d}{2}+\ell-1\right) \Gamma
\left(d+\ell_0-2\right)}{\Gamma\left(\frac{d}{2}+\ell_0-1\right)\Gamma(d+\ell-2)} a_J f^J_{\Delta,\ell,\text{theirs}}$, where $\ell_0$ is the spin of the superconformal primary and $a_{\frac{p}{2}-1}=1$, $a_{\frac{p}{2}}=\frac{p}{1-p}$, $a_{\frac{p}{2}+1}=\frac{(p+2)(p+1)}{p(p-1)}$ for $\langle 2p2p\rangle$ and the same expressions with $p=2$ for $\langle22pp\rangle$. } For instance, the top component is:
\es{top}{
  f^{p/2+1}_{\Delta+2,\ell}=\frac{(\Delta -(p-2) \epsilon +\ell ) (-\Delta +p \epsilon +\ell )}{(\Delta-p \epsilon +\ell +2) (-\Delta +(p+2)\epsilon +\ell -2)}\,.
}
The multiplets that appear in each OPE are
\es{OPEs}{
\mathcal{D}[2]\times \mathcal{D}[p]=\sum_{j=0}^2\mathcal{D}[p+2-2j]+\sum_{j=1}^2\sum_{\ell=0}^\infty\mathcal{B}[p+2-2j]_\ell+\sum_{\ell=0}^\infty\sum_{\Delta}\mathcal{L}[p-2]_{\Delta,\ell}\,,
}
and have the following scaling dimensions:
\es{multDim}{
\Delta_{\cL[p]}>\epsilon p+\ell+\mu\,,\qquad \Delta_{\mathcal{B}[p]}=\epsilon p+\ell+2\epsilon\,,\qquad \Delta_{\mathcal{D}[p]}=\epsilon p\,,
}
where $\mu=2\epsilon$ for $d=3$ and $\mu=4\epsilon-2$ for $d=5,6$. 
For $\langle22pp\rangle$, the identity multiplet $\cD[0]$ and the stress tensor multiplet $\cB[0]_0$ can only appear in the singlet irrep $\bf r=1$, the flavor multiplet $\cD[2]$ can only appear in the adjoint $\bf r=adj$, the $\cD[4]$ and $\cL[0]$ multiplets can only appear in irreps in the symmetric product of two adjoints, while $\cB[2]_\ell$ can only appear in the antisymmetric product. Also, $\cB[0]_\ell$ for $\ell>0$ correspond to spin $\ell+2$ higher spin conserved currents that only exist for free theories. We normalize the OPE coefficients of the identity and the conserved currents as \cite{Chang:2017xmr}
\es{identity}{
\lambda_{pp \cD[0]}^2=1\,,\qquad \lambda_{pp\cB[0]_0}^2= \left(\frac{p}{2}\right)^2\frac{4(2\epsilon+2)(2\epsilon+3)}{2\epsilon+1}\frac{1}{c_T}\,,\qquad \lambda^2_{pp\cD[2]}=\frac{2\epsilon+1}{\epsilon}\frac{2h^\vee}{c_J}\,,
}
where $h^\vee$ is the dual Coxeter number, the central charges $c_T$ and $c_J$ were defined in \eqref{charges}, and the middle equation is consistent with \eqref{cTrel}.

We can in fact obtain all the superblocks of short multiplets by taking limits of long multiplet superblocks as~\cite{Chang:2017xmr}:
 \es{longToShort}{
 \mathcal{D}[p-2] ={}& \lim_{\Delta\to(p-2)\epsilon}\mathcal{L}[p-2]_{\Delta,0}\,,\\
 \mathcal{D}[p]={}&\lim_{\Delta\to p\epsilon-1}\frac{\Delta -p \epsilon +1}{\Delta-(p-2) \epsilon -1}\mathcal{L}[p-2]_{\Delta,-1}\,,\\
\mathcal{D}[p+2]={}&\lim_{\Delta\to (p+2)\epsilon-2}\frac{(\Delta -p \epsilon +2) (\Delta
   -(p+2) \epsilon +2)}{(\Delta -(p-2)
   \epsilon ) (\Delta -p \epsilon )}\mathcal{L}[p-2]_{\Delta,0}\,,\\
\mathcal{B}[p-2]_\ell={}&\lim_{\Delta\to p\epsilon+\ell} \mathcal{L}[p-2]_{\Delta,\ell}\,,\\
\mathcal{B}[p]_\ell={}& \lim_{\Delta\to (p+2)\epsilon+\ell-1}\frac{(\ell+\epsilon ) (\ell+\epsilon +1)(-\Delta +\ell+(p+2) \epsilon -1)}{(\ell+1)(\ell+2 \epsilon ) (-\Delta +\ell+p \epsilon+1)}\mathcal{L}[p-2]_{\Delta,\ell+1}\,.
 }
Note that for $\mathcal{D}[p]$ the limit involves setting the spin of the long superblock superprimary to negative one.

\subsection{Reduced correlator}
\label{reduced}

Instead of solving the Ward identities by expanding in superblocks, we can try to find an exact solution. In position space, such a formal solution was found in \cite{Dolan:2004mu}, and takes the form
\es{redPos}{
G_{\bf r}(U,V,\alpha)=&U^{2\epsilon}\Delta_{\epsilon}\left[(V+U (\alpha-1)-1)\alpha+1\right]\cG_{\bf r}(U,V)\,,
}
where the differential operator $\Delta_\epsilon$ is defined as
\es{Delta}{
&D_\epsilon=U{\partial^2\over\partial U^2}+V{\partial^2\over\partial V^2} +(U+V-1){\partial^2\over \partial U\partial V}+(1+\epsilon)\left({\partial\over \partial U}+{\partial\over \partial V}\right)\,,
\\
&\Delta_\epsilon=(D_\epsilon)^{\epsilon-1}U^{\epsilon-1}\,.
}
The reduced correlator $\cG_{\bf r}(U,V)$ now solves the Ward identities for any function of $U,V$, and also does not depend on $\alpha$. The problem is that $\Delta_\epsilon$ is only well defined for even $d$, as otherwise it contains fractional derivatives. In these cases, a superblock expansion for $\cG_{\bf r}(U,V)$ was given in $d=4$ \cite{Dolan:2001tt,Beem:2014zpa} and $d=6$ \cite{Chang:2017xmr}, and consists of just a single conformal block for each supermultiplet.

We will avoid the issue of fractional derivatives by working in Mellin space, following the approach of \cite{Virally:2025nnl}. We define the Mellin transforms of $G$ and $\widetilde{G}$ as
\es{mellin}{
{G}_\text{conn}^{ABCD}(U,V,\alpha)&=\int \frac{ds dt}{(4\pi i)^2}U^{\frac s2}V^{\frac{t-\epsilon(p+2)}{2}} M^{ABCD}(s,t,\alpha)\\
&\hspace{25mm} \times \Gamma\Big[2\epsilon-\frac s2\Big]\Gamma\Big[p\epsilon-\frac s2\Big]\Gamma\Big[\frac{(2+p)\epsilon-t}{2}\Big]^2\Gamma\Big[\frac{(2+p)\epsilon-u}{2}\Big]^2\,,\\
\widetilde{{G}}_\text{conn}^{ABCD}(U,V,\alpha)&=\int \frac{ds dt}{(4\pi i)^2}U^{\frac {s}{2}}V^{\frac{t-\epsilon(p+2)}{2}} \widetilde{M}^{ABCD}(s,t,\alpha)\\
& \hspace{25mm} \times\Gamma\Big[\frac {(2+p)\epsilon-s}{2}\Big]^2\Gamma\Big[\frac{(2+p)\epsilon-t}{2}\Big]^2\Gamma\Big[2\epsilon-\frac{u}{2}\Big]\Gamma\Big[p\epsilon-\frac{u}{2}\Big]\,,\\
}
where $s+t+{u}=2\epsilon(2+p)$, the two integration contours include all poles in $s,t$ but not $u$, and we define the connected correlator by subtracting the disconnected correlator as 
\es{connected}{
{G}_\text{conn}^{ABCD}(U,V,\alpha)&={G}^{ABCD}(U,V,\alpha)-\left(\delta^{AB}\delta^{CD}+\delta_{2,p}U^{2\epsilon}\alpha^2\delta^{AC}\delta^{BD}+\delta_{2,p}\frac{U^{2\epsilon}}{V^{2\epsilon}}(1-\alpha)^2\delta^{AD}\delta^{BC}\right)\,,\\
\widetilde{{G}}_\text{conn}^{ABCD}(U,V,\alpha)&=\widetilde{{G}}^{ABCD}(U,V,\alpha)-\left(\delta_{2,p}\delta^{AB}\delta^{CD}+U^{\epsilon\frac{p+2}{2}}\alpha^{\frac{p+2}{2}}\delta^{AC}\delta^{BD}+\delta_{2,p}\frac{U^\epsilon}{V^{2\epsilon}}(1-\alpha)^2\delta^{AD}\delta^{BC}\right)\,.
}
The Mellin amplitudes then transform under crossing as
\es{melCross}{
M^{ABCD}(s,t,\alpha)=M^{BACD}(s,u,1-\alpha)=\alpha^{\frac{p+2}{2}}\widetilde{M}^{ACBD}(u,t,1/\alpha)\,,
}
which can be derived from the position space crossing in \eqref{cross}.

In Appendix \ref{blockDets}, we show how the position space Ward identities \eqref{ward} can be converted to Mellin space and solved in terms of a reduced Mellin amplitude $\cM^{ABCD}(s,t)$ as
\es{redM}{
M^{ABCD}(s,t,\alpha)&=\sum_{J=0}^{2}{\mathcal{Y}}^2_J(\alpha) \cD^\epsilon_J(s,t)\cM^{ABCD}(s,t)\,,\\
\widetilde{M}^{ABCD}(s,t,\alpha)&=\sum_{J=p/2-1}^{p/2+1}{\mathcal{Y}}^p_J(\alpha) \widetilde{\cD}^\epsilon_J(s,t)\widetilde{\cM}^{ABCD}(s,t)\,,\\
}
where $\cD^\epsilon_J$ are polynomials in $s,t$ multiplying shifts in $s,t$ acting on $\cM^{ABCD}(s,t)$, whose explicit form is given in Appendix \ref{blockDets}. The reduced Mellin amplitudes transform under crossing as\footnote{When going from the $\langle 22pp\rangle$ correlator to the $\langle 2p2p\rangle$ correlator or vice-versa using the crossing relations in reduced form, we must then use the correct operator, $\mathcal{D}_J^\epsilon$ or $\widetilde{\mathcal{D}}_J^\epsilon$, to return to the full expression.}
\es{melCross2}{
\cM(s,t)=\cM(s,\tilde u)=\widetilde{\cM}(\tilde u,t)\,,
}
where we define $s+t+\tilde u=2( p\epsilon+2\epsilon-2)$. 

We can consider an analogue of the block expansion in Mellin space, that is useful for the strong coupling expansion we consider in the next section. While it is complicated to write a single conformal block $g^{\Delta_{12},\Delta_{34}}_{\Delta,\ell}(U,V)$ in Mellin space, one can define a function that has the same poles in $s$ as the block would have \cite{Mack:2009mi,Mack:1975jr}:
\es{mellinBlock}{
M^{\Delta_{12},\Delta_{34}}_{\Delta,\ell}(s,t)=\sum_{m=0}^\infty\frac{K_{\Delta,\ell,m}^{\Delta_1,\Delta_2,\Delta_3,\Delta_4}Q^{\Delta_{12},\Delta_{34}}_{\Delta,\ell,m}(u-\Delta_1-
\Delta_4)}{s-(\Delta-\ell+2m)}\,,
}
where explicit expressions for the prefactor $K$ and the Mack polynomial $Q$ are reviewed in Appendix \ref{blockDets}. This expression corresponds in AdS$_{d+1}$ to a Witten exchange diagram for an operator with the scaling dimension $\Delta$ and spin $\ell$. Note that the poles in $s$ correspond to the twist of the exchanged operator, where the sum over $m$ corresponds to the conformal descendants. One can then check if the combination of exchange diagrams corresponding to a long superblock \eqref{superExp} in the full correlator can be written in terms of a reduced Mellin amplitude $\cM_{\Delta,\ell}$ using the difference operator in \eqref{redM}:
\es{redBlock}{
&\sum_{a=0}^4\sum_{b=-2}^2f^J_{\Delta+a,\ell+b}M^{0,0}_{\Delta+a,\ell+b}(s,t)\approx {\cD}_J^\epsilon(s,t)\cM_{\Delta,\ell}(s,t)\,,\\
&\sum_{a=0}^4\sum_{b=-2}^2 f^J_{\Delta+a,\ell+b}M^{\epsilon(2-p),\epsilon(2-p)}_{\Delta+a,\ell+b}(s,t)\approx \widetilde{\cD}_J^\epsilon(s,t)\widetilde{\cM}_{\Delta,\ell}(s,t)\,,\\
}
where $\approx$ means that we ignore terms that are entire in $s$, and note that these equations must be satisfied for all three values of $J$. In both cases we find a solution in terms of a single exchange diagram
\es{redBlock2}{
\cM_{\Delta,\ell}(s,t)&=\sum_{m=0}\frac{-f^2_{\Delta+2,\ell} K_{\Delta+2,\ell,m}^{\epsilon p+1,\epsilon p+1,d-2,d-2} Q_{\Delta+2,\ell,m}^{2,0}(\epsilon(p+2)-4-s-t)}{2(\Delta+2+\ell)(\Delta+2-\ell-d)(s-(\Delta-\ell)-2m)}\,,\\
\widetilde{\cM}_{\Delta,\ell}(s,t)&=\sum_{m=0}\frac{-f^{\frac p2-1}_{\Delta+2,\ell} K_{\Delta+2,\ell,m}^{d-2 ,\epsilon p ,d-2,\epsilon p+2} Q_{\Delta+2,\ell,m}^{2+\epsilon(p-2),\epsilon(p-2)}(\epsilon(p+2)-4-s-t)}{2(\Delta-\epsilon p-\ell)(\Delta-(p-2)\epsilon+\ell)(s-(\Delta-\ell)-2m)}\,,\\
}
where the top component $f$ is defined in \eqref{top}, and it can be shown that each of these expressions is equivalent when $p=2$. Note that the right hand side almost resembles a standard exchange diagram \eqref{mellinBlock}, except $\Delta$ in the pole in $s$ is shifted relative to $K$ and $Q$ by two, and the superscripts of the latter do not align. 

We can then obtain similar expressions for the short supermultiplets by taking the limits in \eqref{longToShort}. The exception is the $\cD[2]$ multiplet in $\langle22pp\rangle$ and the $\cD[p]$ multiplet in $\langle 2p2p\rangle$, since the Mack polynomials are not defined for negative spin. In even $d$, this could be avoided by taking the reduced correlator in position space for this superblock, which is defined for negative spin, and explicitly taking the Mellin transform. But for odd $d$ this cannot work, which shows the limitation of the reduced correlator in this case, but thankfully will not affect the graviton exchange diagram that we consider in the next section.
	
\section{Gluon scattering in {\normalfont AdS}$_{d+1}\times S^3$}\label{sec:scattering}

We will now study the correlators $\langle 22pp\rangle$ for $d=3,6$ in the large $N$ expansion, where they are holographically dual to gluon (and higher KK mode) scattering on AdS$_{d+1}\times S^3$ in M-theory. For each $d$, we will start by reviewing the general form of the large $N$ expansion as fixed from the analytic bootstrap, including the leading gluon exchange diagram from \cite{Alday:2021odx}. We will then use the cubic couplings from Section \ref{sec:sugra} to compute the graviton exchange term, which takes an especially simple form when written in terms of the reduced Mellin amplitude. We confirm our results by taking the flat space limit and comparing to the known flat space amplitudes. For the 6d case, we also perform the unmixing of the single and double trace operators.

\subsection{AdS$_{4}\times S^3$}
\label{ads4}

We start by considering the $d=3$ case, where $\langle 22pp\rangle$ describes gluons scattering in M-theory on AdS$_4\times S^7/\mathbb{Z}_{N_f}$, which contains a $\mathbb{C}^2/\mathbb{Z}_{N_f}$ orbifold singularity, with a fixed point locus of AdS$_4\times S^3$. The flavor symmetry in this case includes $G_F=SU({N_f})$. The analytic bootstrap restricts the large $N$ expansion to be \eqref{largeN3d} \cite{Chester:2023qwo,Alday:2021odx}, where $c_J$ and $c_T$ are defined by \eqref{charges} and given in Table \ref{Tab: relevant constants}. The leading term $M_F^2$ corresponds to a tree level gluon exchange diagram, and takes the explicit form \cite{Alday:2021odx}\footnote{In the notation of that paper, we use the longest root $\psi^2=2$, the dual Coxeter number $h^\vee= {N_f} $, and the relation $\texttt{c}^{ABCD}_s=\psi^2 h^\vee P_{\bf adj}^{ABCD}$.}
\es{leading3d}{
&M_{F^2}^{ABCD}=16 {N_f} P_{\bf adj}^{ABCD}\frac{(2 \alpha  (-p+t+u-2)+p-2 u+2) \, _3F_2\left(\frac{1}{2},\frac{3-p}{2},\frac{1-s}{2};\frac{3}{2},\frac{3-s}{2};1\right)}{\pi ^{5/2} (s-1) \Gamma \left[\frac{p-1}{2}\right]}\\
&+16 {N_f} P_{\bf adj}^{ADBC}\frac{(1-\alpha ) \Gamma \left[\frac{p}{2}\right] (2 \alpha  (s-p)-p+2 u-2) \, _3F_2\left(\frac{1}{2},\frac{1}{2},\frac{p-2t}{4};\frac{p+1}{2},\frac{p-2t+4}{4};1\right)}{\pi ^2 (2 t-p) \Gamma \left[\frac{p-1}{2}\right] \Gamma \left[\frac{p+1}{2}\right]}\\
&-16 {N_f} P_{\bf adj}^{ACDB}\frac{\alpha\,  \Gamma \left[\frac{p}{2}\right] (2 \alpha  p-3 p-2 (\alpha -1) s+2 t-2) \, _3F_2\left(\frac{1}{2},\frac{1}{2},\frac{p-2u}{4};\frac{p+1}{2},\frac{p-2u+4}{4};1\right)}{\pi ^2 (2 u-p) \Gamma \left[\frac{p-1}{2}\right] \Gamma \left[\frac{p+1}{2}\right]}\,,
}
which has poles in $s,t,u$ corresponding to the single trace operators in the flavor multiplet $\cD[2]$ with twist one and all their infinite descendants, as well as the cross channel multiplet $\cD[p]$ with twist $p/2$ and its infinite descendents. At large $s,t$ it becomes
\es{gluon3dLarge}{
M_{F^2}^{ABCD}\sim \frac{16N_f(u+s\alpha)^2}{\pi^2\Gamma(p/2)}\left[\frac{P_\text{\bf adj}^{ADBC}}{tu}-\frac{P_\text{\bf adj}^{ABCD}}{su}\right]\,.
}
None of the other terms shown in \eqref{largeN3d} have been computed yet. The 1-loop $M_{F^2|F^2}$ gluon exchange can be fixed from tree level data using the AdS unitarity method \cite{Aharony:2016dwx}, and would have poles for all double trace operators with integer twists above one. The 2-loop $M_{F^2|F^2|F^2}$ gluon exchange could in principle be fixed from tree and 1-loop data as discussed in the analogous 4d case in \cite{Huang:2023oxf}, and would have poles for both double and triple trace operators with integer twists above one. The $M_{D^2F^4,i}$ are contact terms from higher derivative corrections $D^2F^4$, and are polynomials in $s,t$, whose coefficients $\kappa_i$ require inputs beyond the analytic bootstrap.

We will focus here on computing the tree level graviton exchange term $M_R$. In the direct channel, it gets contributions from scalars with scaling dimensions $k_2/2=1,2,3,\dots$, where $k_2/2=1$ corresponds to the stress tensor multiplet $\cB[0]_0$ and $k_2/2>1$ corresponds to long multiplets in the singlet irrep of the flavor symmetry. The crossed channels gets contributions from $k_p/2=p/2,p/2+1,p/2+2,\dots$, where again $k_p/2=p/2$ is a protected multiplet $\cB[p-2]_0$ and $k_p/2>p$ are long multiplets. We can thus write the reduced correlator graviton exchange as
\es{grav3d}{
\cM^{ABCD}_{R}(s,t)&=c_T\sum_{k_2}\lambda_{22k_2}\lambda_{ppk_2}\cM_{k_2,0}(s,t)\delta^{AB}\delta^{CD}\\
&+c_T\sum_{k_p}\lambda_{p2k_p}^2 \Big(\widetilde{\cM}_{k_p,0}(\tilde u,t)\delta^{AC}\delta^{BD}+\widetilde{\cM}_{k_p,0}(t,\tilde u)\delta^{AD}\delta^{BC}\Big)\,,
}
where we used the crossing equations \eqref{melCross2} to relate the cross channels, and the factor of $c_T$ is because we defined $M_R$ to multiply $1/c_T$ in \eqref{largeN3d}. Naively, the expressions for the scalar exchange diagrams in \eqref{redBlock2} are only nonzero for $k_2=2$ and $k_p=p$, while the OPE coefficients $\lambda_{pqk_r}^{(d=3)}$ in \eqref{lam3d} diverge for all the other values of $k_r$. The resolution is that the scalar exchange diagrams should be written directly in terms of the cubic couplings \eqref{Eq: cubic coupling general d n}, which are finite and nonzero for all values, which in practice is the same as cancelling the zero from \eqref{redBlock2} with the infinity from \eqref{lam3d}. We thus find
\es{grav3d2}{
\cM_{R}&=-\sum_{\Delta=1,m=0}^\infty\frac{\delta^{AB}\delta^{CD}}{s-2m-\Delta}
\frac{(2 \Delta +1)  \Gamma \left(\frac{\Delta +3}{2}\right) \Gamma
   \left(m+\frac{\Delta }{2}+1\right) \Gamma \left(\frac{1}{2} (p+\Delta
   +1)\right) \Gamma \left(m-\frac{p}{2}+\frac{\Delta }{2}+1\right)}{ \pi
   ^{5/2} 2^{-p-3}\Gamma \left(\frac{\Delta }{2}\right) \Gamma (m+1) \Gamma (p-1)
   \Gamma \left(m+\Delta +\frac{3}{2}\right) \Gamma \left(\frac{1}{2}
   (\Delta +2-p)\right)}
\\
&-\sum_{\Delta=p,m=0}^\infty \frac{(2 \Delta +1)  \Gamma \left(\frac{1}{4} (p+2 \Delta
   +4)\right)^2 \Gamma \left(m-\frac{p}{4}+\frac{\Delta
   }{2}+\frac{1}{2}\right) \Gamma \left(m-\frac{p}{4}+\frac{\Delta
   }{2}+\frac{3}{2}\right)}{ \pi ^{5/2}2^{-p-3} \Gamma (m+1) \Gamma (p-1) \Gamma
   \left(m+\Delta +\frac{3}{2}\right) \Gamma \left(\frac{1}{4} (-p+2 \Delta
   +2)\right)^2}\\
   &\qquad\qquad\qquad\times\left[\frac{\delta^{AC}\delta^{BD}}{\tilde u-2m-\Delta}+\frac{\delta^{AD}\delta^{BC}}{ t-2m-\Delta}\right]\,.
}
We can then perform the double sums by first converting to a single sum by considering the finite number of contributions to a given pole in $s,t,\tilde u$, and then doing the single sum, where here and in the following we only consider poles (not entire terms). We find
\es{grav3d3}{
&\cM_{R}=-\delta^{AB}\delta^{CD}\Bigg[\sum_{n=1}^\infty \frac{16 (p-1) p \cos \left(\frac{\pi  p}{2}\right) \Gamma
   \left(n-\frac{p}{2}-\frac{1}{2}\right) \left(\,
   _3F_2\left(1,n-\frac{3}{2},n-\frac{p}{2}-\frac{1}{2};n,n+\frac{1}{2};1\right)-1\right)}{ \pi ^{7/2} (2 n-3) (-2 n+s+1) \Gamma (n)}\\
   &\qquad\qquad\qquad\qquad+ \frac{8 \left(p^2-1\right) s H_{-s}}{ \pi ^2 \Gamma
   \left(\frac{p}{2}\right)}
   \Bigg]+\delta^{AC}\delta^{BD}\Bigg[\frac{4 \left(p^2-1\right) (p-2 (\tilde u+1)) H_{\frac{p}{2}-\tilde u-1}}{ \pi ^2 \Gamma
   \left(\frac{p}{2}\right)}+\\
   &\sum_{n=1}^\infty\frac{16 p \Gamma \left(n-\frac{3}{2}\right) \Gamma
   \left(n+\frac{1}{2}\right) \Gamma \left(\frac{p}{2}+1\right) \left(\,
   _3F_2\left(1,n-\frac{3}{2},n-\frac{3}{2};n,n+\frac{p}{2}-\frac{1}{2};1\right)-1\right)}{ \pi ^3 (2 n-3) \Gamma (n) \Gamma
   \left(\frac{p-1}{2}\right) \left(-2 n-\frac{p}{2}+s+2\right) \Gamma
   \left(n+\frac{p}{2}-\frac{1}{2}\right)}\Bigg]+\text{crossed}\,,
}
where the crossed term is obtained from the $\delta^{AC}\delta^{BD}$ term by swapping $C,D$ and $t,\tilde u$. Note that for odd $p$, only the first $(p+1)/2$ terms in the sum over $n$ are nonzero, due to the terms $\cos(\pi p/2)\Gamma(n-p/2-1/2)$. The hypergeometric terms are subleading in the large $s,t$ limit, where we get
\es{largeMR3d}{
\cM_R^{ABCD}\sim  -\frac{8 \left(p^2-1\right) }{ \pi ^2 \Gamma
   \left(\frac{p}{2}\right)}\left[ \delta^{AB}\delta^{CD}s \log (-s)+\delta^{AC}\delta^{BD}\tilde u \log (-\tilde u)+\delta^{AD}\delta^{BC}t \log (-t) \right]\,.
}
We can check our answer \eqref{grav3d3} by comparing to the flat space M-theory amplitude $\cA(s,t)$ using the flat space limit formula \cite{Penedones:2010ue} for $\langle22pp\rangle$:
    \es{flatGen}{
\mathcal{A}^{ABCD}(s,t)=\lim_{L\to\infty}\frac{{64}\pi^4L^3}{(u+s\alpha)^2}\Gamma\big(\frac{p}{2}\big)\Gamma\big(\frac{p-1}{2}\big) \int\frac{d\beta}{2\pi i}{e^\beta}{\beta^{\frac{1-p}{2}}}M^{ABCD}\Big(\frac{L^2}{2\beta}s,\frac{L^2}{2\beta}t\Big)\,,
}
where in flat space we have $s+t+u=0$. We normalized this formula so that applying \eqref{flatGen} to \eqref{gluon3dLarge} and using the AdS/CFT dictionary in Table \ref{Tab: relevant constants} gives the flat space gluon exchange as computed in \cite{Chester:2023qwo}\footnote{In the notation of \cite{Chester:2023qwo}, we use the relation $\texttt{B}^{ABCD}-\texttt{B}^{ACDB}=\frac{{N_f}}{4}P_\text{\bf adj}^{ABCD}$ and removed the overall polarization factors as well as an overall factor of $4$.}:
 \es{Atree3d}{
\mathcal{A}_{F^2}(s,t)=2 N_f g_\text{YM}^2\left[\frac{P_\text{\bf adj}^{ADBC}}{tu}-\frac{P_\text{\bf adj}^{ABCD}}{su}\right]\,.
}
We can apply the flat space limit to \eqref{largeMR3d} using the large $s,t$ relation
\es{large3dred}{
M(s,t)\sim ((p+1)+2t\partial_t+2s\partial_s)\cM(s,t)\,,
}
and the AdS/CFT dictionary in Table \ref{Tab: relevant constants} to match the flat space graviton amplitude \cite{Chester:2023qwo}\footnote{In the notation of \cite{Chester:2023qwo}, we use the relation $\texttt{A}^{ABCD}=\frac{1}{4}\delta^{AB}\delta^{CD}$ and removed the overall polarization factors as well as an overall factor of $4$.}
 \es{Agrav3d}{
\mathcal{A}_{R}(s,t)&=\frac{\kappa^2 N_f}{2}\int\frac{d^4p_\perp}{(2\pi)^4}\left[\delta^{AB}\delta^{CD}\frac{1}{p_\perp^2-s}+\delta^{AC}\delta^{BD}\frac{1}{p_\perp^2-u}+\delta^{AD}\delta^{BC}\frac{1}{p_\perp^2-t}\right]\\
&\approx-\frac{\kappa^2 N_f}{32\pi^2}\left[\delta^{AB}\delta^{CD}s\log (-s)+\delta^{AC}\delta^{BD}u\log (-u)+\delta^{AD}\delta^{BC}t\log(- t)\right]\,,
}
where the $\approx$ denotes that we only consider logarithmic terms when we integrate over the four transverse momenta between the seven-dimensional orbifold singularity and the ambient 11 dimensions.

We will not perform the unmixing of the single and double trace operators for this theory, as there are other unknown terms in the large $N$ expansion that are more leading.

\subsection{AdS$_{7}\times S^3$}
\label{ads7}

 We next consider the $d=6$ case, where $\langle 22pp\rangle$ describes gluons scattering in M-theory on AdS$_7\times S^4/\mathbb{Z}_2$ in the presence of an M9 brane, with a fixed point locus of AdS$_7\times S^3$. The flavor symmetry in this case includes $G_F=E_8$. The analytic bootstrap restricts the large $N$ expansion to be \eqref{largeN6d}, where $c_J$ and $c_T$ are given in Table \ref{Tab: relevant constants}. The leading term $M_F^2$ corresponds to a tree level gluon exchange diagram, and takes the explicit form \cite{Alday:2021odx}\footnote{In the notation of \cite{Alday:2021odx}, we use the longest root $\psi^2=2$, the dual Coxeter number $h^\vee=30$, and the relation $\texttt{c}^{ABCD}_s=\psi^2 h^\vee P_{\bf adj}^{ABCD}$.}
\es{leading6d}{
&M_{F^2}^{ABCD}=-300P_{\bf adj}^{ABCD}\frac{2 (p (s-4)-3) (2(2 \alpha -1) p-\alpha  (t+u-8)+u-4)}{(2p-3)!\,(s-6)\, (s-4) }\\
&+300P_{\bf adj}^{ADBC}\frac{2 (1-\alpha ) (p (-2 p+t-2)+1) (-2 (2 \alpha +1) p+\alpha  s+u-4)}{(2p-3)!\,(t-2p)\, (t-2 p-2) }\\
&-300P_{\bf adj}^{ACDB}\frac{2 \alpha  (p (-2 k+u-2)+1) (2(2 \alpha -3) p-\alpha  s+s+t-4)}{(2p-3)!\,(u-2p) \,(u-2 p-2)}\,,
}
which has poles in $s,t,u$ corresponding to the single trace operators in the flavor multiplet $\cD[2]$ with twists $4,6$, as well as the cross channel multiplet $\cD[p]$ with twists $2p,2p+2$. The amplitude simplifies in the reduced correlator \eqref{redM}, where it takes the form
\es{leading6dred}{
\cM_{F^2}=\frac{150P_{\bf adj}^{ABCD} (t-2 p)-150P_{\bf adj}^{ADBC} (s-4)}{ (s-4) (2 p-3)! (t-2 p) (-2
   p+s+t-4)}\,,
}
which at large $s,t$ becomes
\es{gluon6dLarge}{
\cM_{F^2}^{ABCD}\sim \frac{150}{(2p-3)!}\left[\frac{P_\text{\bf adj}^{ADBC}}{tu}-\frac{P_\text{\bf adj}^{ABCD}}{su}\right]\,.
}
The calculation of the tree level graviton exchange term $M_R$ is similar to the 3d case considered above. The amplitude gets contributions in the direct channel from scalars with scaling dimensions $2k_2=4,8,12,\dots$, where $2k_2=4$ corresponds to the stress tensor multiplet $\cB[0]_0$ and $2k_2>4$ corresponds to long multiplets in the singlet irrep of the flavor symmetry. The crossed channels gets contributions from $2k_p=2p,2p+4,2p+8,\dots$, where again $2k_p=2p$ is a protected multiplet $\cB[p-2]_0$ and $2k_p>2p$ are long multiplets. The amplitude takes the same form \eqref{grav3d} as in 3d, except the expressions for the OPE coefficients \eqref{lam6d} and exchange diagrams \eqref{redBlock2} are now different. After cancelling the zero in the exchange diagram with the divergent OPE coefficient just as in 3d, we get the analogous double sum
\es{grav6d2}{
&\hspace{-.5in}\cM_{R}=\sum_{\Delta=4,m=0}^\infty\frac{-\delta^{AB}\delta^{CD}}{s-2m-\Delta}
\frac{112 (\Delta -1) \Gamma \left(\frac{\Delta +2}{4}\right) \Gamma
   \left(\frac{\Delta +6}{4}\right) \Gamma \left(m+\frac{\Delta }{2}-2\right)
   \Gamma \left(\frac{1}{2} (4 p+\Delta -2)\right) \Gamma \left(m-2
   p+\frac{\Delta }{2}+1\right)}{ \pi  \Gamma \left(\frac{\Delta
   }{4}-1\right) \Gamma \left(\frac{\Delta }{4}\right) \Gamma (m+1) \Gamma (2
   p) \Gamma (2 p-2) \Gamma (m+\Delta ) \Gamma \left(-2 p+\frac{\Delta
   }{2}+1\right)}
\\
&\hspace{-.5in}+\sum_{\Delta=2p,m=0}^\infty \frac{7 (\Delta -1) 2^{2p+1} p \Gamma \left(p+\frac{\Delta }{2}+1\right)
   \Gamma \left(\frac{1}{4} (2 p+\Delta -2)\right) \Gamma \left(\frac{1}{4}
   (2 p+\Delta +2)\right) \Gamma \left(m-p+\frac{\Delta }{2}-1\right) \Gamma
   \left(m-p+\frac{\Delta }{2}\right)}{ \pi  \Gamma (m+1) \Gamma (2 p-2)
   \Gamma (2 p+1) \Gamma (m+\Delta ) \Gamma \left(\frac{1}{2} (-2 p+\Delta
   -2)\right) \Gamma \left(\frac{1}{4} (\Delta -2 p)\right) \Gamma
   \left(\frac{1}{4} (-2 p+\Delta +4)\right)}\\
   &\qquad\qquad\qquad\times\left[\frac{-\delta^{AC}\delta^{BD}}{\tilde u-2m-\Delta}+\frac{-\delta^{AD}\delta^{BC}}{ t-2m-\Delta}\right]\,.
}
We can then perform the double sums by first converting to a single sum by considering the finite number of contributions to a given pole in $s,t,\tilde u$, and then doing the single sum, where again we only consider poles. While we did not find a general $p$ formula, for the lowest couple values of $p$ we have
\es{grav6d3}{
& \cM^{p=2}_{R}=\delta^{AB}\delta^{CD}\frac{315 \sqrt{\pi } (s-6) (s (7 (s-26) s+1528)-4096) \Gamma
   \left(2-\frac{s}{2}\right)}{2048 \Gamma
   \left(\frac{13}{2}-\frac{s}{2}\right)}\\
   &\qquad+\delta^{AC}\delta^{BD}\frac{315 \sqrt{\pi } (\tilde u-6) (\tilde u (7 (s-26) \tilde u+1528)-4096) \Gamma
   \left(2-\frac{\tilde u}{2}\right)}{2048 \Gamma
   \left(\frac{13}{2}-\frac{\tilde u}{2}\right)}+\text{crossed}\,,\\
   &\cM^{p=3}_{R}=\delta^{AB}\delta^{CD}\frac{147 \sqrt{\pi } (6-s) (2^{20}\cdot15 - 8252032 s + 1661840 s^2 - 162100 s^3 + 7700 s^4 - 143 s^5) \Gamma
   \left(2-\frac{s}{2}\right)}{262144 \Gamma
   \left(\frac{17}{2}-\frac{s}{2}\right)}\\
   &-\delta^{AC}\delta^{BD}\frac{147 \sqrt{\pi } (\tilde u-8) (2457600 - 885040 \tilde u + 116884 \tilde u^2 - 6732 \tilde u^3 + 143 \tilde u^4)
   \Gamma \left(3-\frac{\tilde u}{2}\right)}{131072 \Gamma
   \left(\frac{17}{2}-\frac{\tilde u}{2}\right)}+\text{crossed}\,,\\
}
while we give several higher values of $p$ in the attached \texttt{Mathematica} notebook. For general $p$, we find at large $s,t$:
\es{largeMR6d}{
\cM_R^{ABCD}\sim  \frac{7 \sqrt{2} \Gamma \left(2 p+\frac{3}{2}\right)}{2 \Gamma (2
   p) \Gamma (2 p-2)}\left[ \frac{\delta^{AB}\delta^{CD}}{\sqrt{-s}}+\frac{\delta^{AC}\delta^{BD}}{\sqrt{-u}}+\frac{\delta^{AD}\delta^{BC}}{ \sqrt{-t}} \right]\,.
}
We can check our answer \eqref{grav6d3} by comparing to the flat space M-theory amplitude $\cA(s,t)$ using the flat space limit formula \cite{Penedones:2010ue} for $\langle22pp\rangle$:
    \es{flatGen6}{
\mathcal{A}^{ABCD}(s,t)=\lim_{L\to\infty}12\pi^5\Gamma(2p)\Gamma(2p-2)L^2 \int\frac{d\beta}{2\pi i}{e^\beta}{\beta^{{1-2p}}}(2(p+1)+s\partial_s+t\partial_t)\cM^{ABCD}\Big(\frac{L^2}{2\beta}s,\frac{L^2}{2\beta}t\Big)\,,
}
which is normalized so that applying \eqref{flatGen6} to \eqref{gluon6dLarge} and using the AdS/CFT dictionary in Table \ref{Tab: relevant constants} gives the flat space gluon exchange as computed in \cite{Chester:2023qwo}\footnote{While  \cite{Chester:2023qwo} considered half-maximal 7d SYM, the answer is fixed by supersymmetry to be the same in other dimensions. We convert their notation to ours just as in footnote 15.}:
 \es{Atree6d}{
\mathcal{A}_{F^2}(s,t)=60 g_\text{YM}^2\left[\frac{P_\text{\bf adj}^{ADBC}}{tu}-\frac{P_\text{\bf adj}^{ABCD}}{su}\right]\,.
}
We can apply the flat space limit to \eqref{largeMR6d} using the AdS/CFT dictionary in Table \ref{Tab: relevant constants} to get the expected flat space graviton amplitude
 \es{Agrav6d}{
\mathcal{A}_{R}(s,t)&={\kappa^2}\int\frac{d p_\perp}{2\pi}\left[\delta^{AB}\delta^{CD}\frac{1}{p_\perp^2-s}+\delta^{AC}\delta^{BD}\frac{1}{p_\perp^2-u}+\delta^{AD}\delta^{BC}\frac{1}{p_\perp^2-t}\right]\\
&=-\frac{\kappa^2}{2}\left[\delta^{AB}\delta^{CD}(-s)^{-\frac12}+\delta^{AC}\delta^{BD}(-u)^{-\frac12}+\delta^{AD}\delta^{BC}(-t)^{-\frac12}\right]\,,
}
where compared to the 3d case \eqref{Agrav3d} we have $N_f=2$ because the ambient spacetime is AdS$_7\times S^4/\mathbb{Z}_2$, and there is just a single transverse momentum relative to the 10d worldvolume of the M9 brane.

\subsection{Mixing in AdS$_{7}\times S^3$}

Let us finally discuss operator mixing in the 6d theory. For simplicity let us focus on a particular simple subset of operators. Namely, for each $n=2,3,\dots $ consider the following $(n-1)$ gluon double traces\footnote{There are also double traces with $\Delta=4n+2$, featuring an odd number of boxes, but these do not mix with the graviton modes.}
\begin{align}
  \Phi_p = :\! \phi_p \square^{2(p-2)}\phi_p \! :,\qquad p=2,\dots n\,,
\end{align}
all of which have engineering dimension $\Delta=4n$ and $\ell=0$, and we have contracted indices such that they are all inert under $G_F\times SU(2)_L \times SU(2)_R$. This is a shorthand notation; by each of these operators, we mean the superprimary of a long multiplet built from two gluon multiplets with these specified quantum numbers. Beyond the large $N$ limit this $(n-1)$-fold degeneracy in the spectrum is lifted, and in particular one expects a non-trivial mixing between the $\Phi_p$, with mixing coefficients accessible by studying $\langle ppqq \rangle$. The new ingredient here is that the (super-)extremal cubic couplings between the $\phi_p$ and the graviton modes induces an additional mixing between $\Phi_p$ and the graviton single trace $k_2=2n$. We denote the superprimary of this long multiplet $\rho_{2n}$. The mixing matrix $M$ between these modes is then an $n\times n$ real symmetric matrix defined by
\begin{align}
  &\begin{pmatrix}
  	\langle \rho_{2n}(x) \rho_{2n}(y) \rangle 		& \langle \rho_{2n}(x) \Phi_q(y) \rangle 	\\
  	\langle  \Phi_p(x) \,\rho_{2n}(y) \rangle	& \langle \Phi_p(x) \Phi_q(y) \rangle
  \end{pmatrix} = \frac{1}{|x-y|^{8n}}\Big[ 
  \mathds{1}_n - N^{-3/2}M \log\left(|x-y|^2\right) + \cO\!\left(N^{-2}\right)
  \Big]\,,
\end{align}
where $p,q=2,\dots,n$. The leading correction to $\langle \rho_{2n}(x) \rho_{2n}(y) \rangle$ comes from gluon and graviton loops, both of which come at order $N^{-2}$, and so $M_{11}=0$. By viewing $\langle  \Phi_p(x) \,\rho_{2n}(y) \rangle$ and $\langle \Phi_p(x) \Phi_q(y) \rangle$ as appropriate limits of 3- and 4-point functions, respectively, we see that the leading correction to the former comes at order $N^{-3/2}$, and to the latter at order $N^{-2}$. In particular $M_{pq}=0$ for all $p,q=2,\dots,n$, which indeed can be deduced from the structure of the conformal block expansion of $\langle ppqq\rangle $ as explained below. And so the leading mixing effect is between the graviton single trace $\rho_{2n}$ and each gluon double trace $\Phi_p$, with mixing amongst these double traces a subleading effect. This is in contrast to the AdS$_5$ case \cite{Chester:2025wti} in which these two effects enter at the same order.

The leading mixing is thus controlled by the non-zero elements $M_{1p}$ for $p=2,\dots,n$. Let us expand the OPE coefficient of the mixed operator $\Phi_p$ with $\phi_p\phi_p$ as
\begin{align}
  \lambda_{pp\Phi_q} = \delta_{pq}\lambda^{(0)}_{pp\Phi_p} + \cO \!\left(\frac{1}{N^2}\right)\,,
\end{align}
where the leading term $\lambda^{(0)}_{pp\Phi_p}$ is computable in generalised free field theory. Then at leading order we see that the dimensions of only two operators $\cO_\pm$ are lifted, corresponding to the two non-zero eigenvalues of $M$. Letting
\begin{align}
  |M|:= \sqrt{\sum_{p=2}^n (M_{1p})^2}\,,
\end{align}
their dimensions are
\begin{align}
  \Delta_\pm = 4n \pm \frac{1}{N^{3/2}}|M| + \cO\!\left(\frac{1}{N^2}\right)\,,
\end{align}
while both have the same leading OPE coefficients,
\begin{align}
  \lambda_{pp\cO_\pm} = \frac{1}{\sqrt{2}} \frac{M_{1p}}{|M|}\lambda^{(0)}_{pp\Phi_p} + \cO\left(\frac{1}{N^{1/2}}\right)\,.
\end{align}
The zero eigenvalues of $M$ correspond to the remaining $(n-2)$ unmixed operators $\cO_a$, $a=3,\dots n$. They all have dimensions
\begin{align}
  \Delta = 4n + \cO\!\left(\frac{1}{N^2}\right)\,,
\end{align}
and we can choose a basis for them such that their leading OPE coefficients are
\begin{align}
  \lambda_{22\cO_a} = \frac{M_{1a}}{\sqrt{M_{12}^2 + M_{1a}^2}}\lambda^{(0)}_{22\Phi_2}+ \cO\left(\frac{1}{N^{1/2}}\right),\quad \lambda_{bb\cO_a} = -\frac{\delta_{ab}M_{12}}{\sqrt{M_{12}^2 + M_{1a}^2}}\lambda^{(0)}_{bb\Phi_b}+ \cO\left(\frac{1}{N^{1/2}}\right)\,.
\end{align}
We finally need to explain how to determine the leading non-zero mixing coefficients $M_{1p}$ for $p=2,\dots,n$. They are accessible from the graviton exchange contribution to $\langle 22pp\rangle$ by virtue of the superblock expansion (\ref{superblockExp}). If we write the singlet contribution to $\langle 22pp \rangle $ as
\begin{align}
  G_\mathbf{1}(U,V;\alpha) = G^{(0)}_\mathbf{1}(U,V;\alpha) + \frac{1}{N^{3/2}}G^{(3/2)}_\mathbf{1}(U,V;\alpha)  + \frac{1}{N^2}G^{(2)}_\mathbf{1}(U,V;\alpha)  + \frac{1}{N^3}G^{(3)}_\mathbf{1}(U,V;\alpha) + \dots \,,
\end{align}
then the leading data we want is encoded in the leading logarithm terms in a small $U$ expansion of these terms as
\begin{align}
  G_\mathbf{1}^{(0)}(U,V;\alpha)|_{\Delta=4n,\ell=0} 		&= (v_2 \cdot  v_p) \mathfrak{G}_{4n,\ell=0}(U,V,\alpha) = \delta_{2p} (\lambda^{(0)}_{22\Phi_2})^2\mathfrak{G}_{4n,\ell=0}(U,V,\alpha)		\nn\\
  G_\mathbf{1}^{(3/2)}(U,V;\alpha)|_{\Delta=4n,\ell=0} 		&= \frac{1}{2}(\log U) (v_2\cdot M v_p) \mathfrak{G}_{4n,\ell=0}(U,V,\alpha) + \dots 		\nn\\
  G_\mathbf{1}^{(3)}(U,V;\alpha)|_{\Delta=4n,\ell=0} 		&= \frac{1}{8}(\log U)^2 (v_2\cdot M^2 v_p) \mathfrak{G}_{4n,\ell=0}(U,V,\alpha) + \dots \,,
\end{align}
where we have extracted the contribution from operators of engineering dimension $\Delta=4n$ and spin $\ell=0$, and in each expression, the $\dots$ denote terms with a lower power of $(\log U)$ which contain higher order CFT data. Meanwhile $v_p$ is the $n$-component vector
\begin{align}
  (v_p)_1 = 0,\quad (v_p)_q = \delta_{pq}\lambda^{(0)}_{pp\Phi_p}\,.
\end{align}
We have of course that\footnote{Similarly, an absence of an $N^{-3/2}\log U$ term in $\langle ppqq\rangle $ confirms that $M_{pq}=0$.} $ G_\mathbf{1}^{(3/2)}(U,V;\alpha)=0$ identically, which we see just corresponds to the fact that $M_{2p}=0$ for all $p=2,\dots, n$. Meanwhile from $G_\mathbf{1}^{(3)}(U,V;\alpha)$ we extract
\begin{align}
  M_{12}M_{1p} =  \frac{(v_2\cdot M^2 v_p)}{\lambda^{(0)}_{22\Phi_2}\lambda^{(0)}_{pp\Phi_p}}\,,
\end{align}
from which, given the GFFT OPE coefficients $\lambda^{(0)}_{pp\Phi_p}$, we can uniquely determine the $M_{1p}$ up to an overall sign, which just corresponds to our ability to flip the sign $\rho_{2n}\to -\rho_{2n}$.

Let us finally provide some explicit numbers for the lowest dimension case $n=2$, so $\Delta=8$. We find then
\begin{align}
  \lambda^{(0)}_{22\Phi_2} = \sqrt{\frac{2}{3}}\,,
\end{align}
and
\begin{align}
  (v_2\cdot M^2 v_2) = \frac{15 \dim \frak{g}}{14} = \frac{1860}{7} \quad \Longrightarrow \quad M_{12}=\frac{2790}{7}\,,
\end{align}
from which it follows that the two unmixed operators have scaling dimensions
\begin{align}
  \Delta_\pm = 8 \pm \sqrt{\frac{2790}{7}}\frac{1}{N^{3/2}} + \cO\!\left(\frac{1}{N^2}\right)\,,
\end{align}
and OPE coefficients
\begin{align}
  \lambda_{22\cO_\pm} =  \sqrt{\frac{1}{3}}+ \cO\!\left(\frac{1}{N^{1/2}}\right)\,.
\end{align}

\section{Conclusion}\label{sec:conclusion}

In this paper, we computed the bulk couplings $\beta_{pqk_r}$ between gluon KK modes $p,q$ and graviton KK modes $k_r$ for M-theory on AdS$_4\times S^7/\mathbb{Z}_{N_f}$ and AdS$_7\times S^4/\mathbb{Z}_2$. We then used these couplings to compute the graviton exchange correction to the gluon correlators $\langle22pp\rangle$, and matched these to the expected flat space limits. These corrections took a particularly simple form as written in terms of the so-called reduced correlator solution to the superconformal Ward identity, which we derived for CFTs with eight supercharges in general $3\leq d\leq6$. We also performed the unmixing of single and double trace operators in the AdS$_6$ case, which gives the leading large $N$ correction to CFT data beyond the leading gluon exchange term.

Together with our previous work \cite{Chester:2025wti}, we have now considered holographic theories with eight supercharges with extremal couplings in every dimension except $d=5$. The simplest family of $d=5$ holographic CFT are the Seiberg exceptional theories \cite{Seiberg:1996bd}, which are $USp(2N)$ gauge theories coupled to $N_f\leq 7$ hypermultiplets in the fundamental and one hypermultiplet in the antisymmetric. The bulk dual is engineered by $N$ D4 branes and certain configurations of D8 branes and an O8 orientifold in type IIA string theory, which at large $N$ is described by supergravity on a warped product of AdS$_6$ and a hemisphere $HS^4$ \cite{Ferrara:1998gv,Brandhuber:1999np,Bergman:2012kr}. Due to the warping, it is much more challenging to compute the bulk coupling between gravitons and gluons on the D8. It would be particularly interesting to compute the graviton exchange contribution to gluon scattering in this theory, as much like the 6d case considered in this paper, this is the leading correction to gluon exchange at strong coupling, and has no contact term ambiguities.

One of the more general technical advances of this paper was a reduced correlator in Mellin space for $3\leq d\leq6$, which generalized the previously known position space reduced correlator that was only valid in $d=4,6$ \cite{Dolan:2004mu,Chang:2017xmr}. In particular, we showed how exchange diagrams for all superblocks can be written in the reduced correlator language, except for the flavor multiplet in odd $d$. Since the gluon exchange term is related to the flavor multiplet, this is why we were unable to write this term in the reduced correlator format for 3d. This is likely related to the fact that the flavor multiplet is related to the long multiplet by a limit where we take the spin to be negative, which has no obvious interpretation in Mellin space. If one could overcome this problem, then one could perhaps look for an analogous position space reduced correlator in odd $d$, which could drastically simplify numerical bootstrap studies.

In 3d, the graviton exchange correction was not the first correction to the gluon exchange term at strong coupling, unlike the 6d case. Instead, the 1-loop gluon exchange term is more leading. It would thus be interesting to compute this correction following the standard AdS unitarity method \cite{Aharony:2016dwx}, which has already been applied to 3d 1-loop terms with maximal supersymmetry in \cite{Alday:2021ymb,Alday:2022rly}, as well as half-maximal 1-loop terms in 4d \cite{Alday:2021ajh}. Hopefully this term can be written in reduced correlator format, which should be drastically simpler than the very complicated expressions in \cite{Alday:2021ymb,Alday:2022rly}. Since we have such a reduced correlator for general $3\leq d\leq6$, it might even be possible to compute this 1-loop term as an analytic function of $d$, which should interpolate to the known 4d result \cite{Alday:2021ajh,Huang:2023ppy}.\footnote{In the maximally supersymmetric case, 1-loop graviton exchange amplitude have been computed in all possible dimensions: 3d \cite{Alday:2021ymb,Alday:2022rly}, 4d \cite{Alday:2017xua,Aprile:2017bgs,Alday:2019nin,Alday:2021vfb}, and 6d \cite{Alday:2020tgi}.}

Finally, once we have computed the leading corrections to tree gluon exchange for all $3\leq d\leq6$, one could see how these corrections compare to a numerical bootstrap study at large but finite $N$. In particular, following a similar strategy as graviton scattering in maximally supersymmetric theories \cite{Alday:2022ldo}, one can use the fact that corrections to tree gluon exchange from either gravitons or 1-loop terms are sensitive to the precise KK spectrum, which allows one to distinguish between the known holographic theories considered here, and putative pure holographic theories with no higher KK modes. One can then check which theory saturates the bound, and in particular if the pure theories saturate the bound as was shown for the maximally supersymmetric case in \cite{Alday:2022ldo}.

\section*{Acknowledgments} 
We thank Ofer Aharony, Fernando Alday, Pietro Ferrero, Silviu Pufu, Yifan Wang, Xi Yin, and Deliang Zhong for useful discussions. SMC is supported by the Royal Society under the grant URF\textbackslash R1\textbackslash 221310 and the UK Engineering and Physical Sciences Research council grant number EP/Z000106/1, RM is also supported by EP/Z000106/1, JvM is supported by the STFC Consolidated Grant ST/X000575/1.  JvM would like to thank the ITF at the KUL for their continuous hospitality. The work of CV is supported by the Fonds de recherche du Qu\'ebec, secteur Nature et technologies, through its Doctoral Training Scholarships program, scholarship number 349150 (\hyperlink{https://doi.org/10.69777/349150}{https://doi.org/10.69777/349150}).

\appendix \addtocontents{toc}{\protect\setcounter{tocdepth}{1}}

\section{Details of the computation of cubic couplings }\label{app: cubic couplings}

This Appendix provides supplementary material for the calculation of bulk cubic couplings, summarised in Section \ref{subsec: cubic couplings}.

\subsection{Diagonalisation of supergravity fluctuations}\label{subsec: diag}

The first step is to diagonalise the scalar fluctuations arising from the background metric and form fields, for which we can rely on the results in \cite{vanNieuwenhuizen:1984iz,Castellani:1984vv,Bastianelli:2000mk,Bastianelli:1999en}. The metric fluctuations can be decomposed into scalars, vectors, and a symmetric tensor. Gauge fixing these fluctuations we have that the relevant variations take the form\footnote{The $h$-dependent term in $\delta g_{\mu\nu}$ is a standard Weyl transformation that is imposed to simplify the Einstein equations.}
\begin{equation}\label{Eq: fluctuations}
	\delta g_{\mu\nu} = h'_{\mu\nu} + \frac{h'}{d+1} g_{\mu\nu} - \frac{h}{d-1} g_{\mu\nu}\,,\quad \delta g_{\alpha\beta} = h'_{\alpha \beta} + \frac{h}{n} g_{\alpha\beta}\,,\quad g^{\alpha\beta} h'_{\alpha\beta} = g^{\mu\nu} h'_{\mu\nu} = 0\,,
\end{equation}
where the $h$-fields can in turn be expanded into harmonics on the internal geometry. Merely keeping the scalar harmonics these expansions are
\begin{equation}
	h'_{\mu\nu} = \sum_{I_k} h_{\mu\nu}^{I_k}(x)\, \mathcal X^{I_k}(y)\,,\quad h' = \sum_{I_k} h'^{I_k}(x) \mathcal X^{I_k}(y)\,,\quad h = \sum_{I_k} h^{I_k}(x) \mathcal X^{I_k}(y)\,,
\end{equation}
where
\begin{equation}
	\square_{S} \mathcal X^{I_k} = - \epsilon^2 k(k + n - 1) \mathcal X^{I_k}\,.
\end{equation}
In Section \ref{subsec: cubic couplings} we provide relevant details on these harmonics, and their vector counter parts. The eleven-dimensional Einstein equations along the AdS directions relate the two scalar fluctuations such that
\begin{equation}
	h'^{I_k}(x) = \frac{2}{n} \frac{n + d - 1}{d - 1} h^{I_k}(x)\,.
\end{equation}
After gauge fixing the fluctuations of the gauge potential $C_{n-1}$ we have that the only surviving component relevant for our analysis takes the form
\begin{equation}
	\delta C_{n-1} = \star_{S^{n}} (\rmd a)_{S^n}\,,\quad \text{where} \quad a = \sum_{I_k} a^{I_k}(x) \mathcal X^{I_k}(y)\,.
\label{eq: C n-1 variation}
\end{equation}
Plugging in the fluctuations into the Einstein equations one finds a system of coupled scalars whose diagonalised form equals
\begin{equation}
	\square_{\text{AdS}} s^{I_k} = \epsilon^2 k(k-n+1) s^{I_k}\,,\quad \square_{\text{AdS}} t^{I_k} = \epsilon^2 (k+n-1)(k+2n-2) t^{I_k}\,,
\end{equation}
where
\begin{equation}
	h^{I_k} = \frac{n}{3} (k s^{I_k} + (k+n-1)t^{I_k})\,,\quad a^{I_k} = -\frac{2(n-1)(d+n-1)}{6d^2}(s^{I_k} - t^{I_k})\,.
\end{equation}
Holographically, the scalars $s^{I_k}$ and $t^{I_k}$ correspond to superconformal primaries and their descendents, with conformal dimensions
\begin{equation}
	\Delta_{s^{I_k}} = \epsilon k\,,\quad \Delta_{t^{I_k}} = 2d+\epsilon k\, .
\end{equation}
Since we are only interested in the superconformal primary operators we will from now on drop all $t^{I_k}$ dependences. The associated quadratic action of the scalar modes equals
\begin{equation}
	 S_{s^{I_k}} = b(k) \epsilon^{-n} \int \rmd^{d+1} x \sqrt{-g_{d+1}}\left( -\frac12 s^{I_k} \square s^{I_k} - \frac12 m_{I_k}^2 (s^{I_k})^2 \right) \int \rmd^n y \sqrt{g_{S^n}} (\mathcal X^{I_k})^2\,,
\end{equation}
where
\begin{equation}
	b(k) = \frac{1}{2\kappa^2}\frac{9 k(k-1)(2k+n-1)}{d(dk + n - 1)} \,,
\end{equation}
is the same function appearing in (\ref{eq: SUGRA kinetic term}) and $\sqrt{g_{S^n}}$ is the volume form on the unit round $S^n$.

So far, at quadratic order, we merely had to diagonalise the fluctuations coming from $h$ and $a$. At higher order in the fluctuations however there is a mixing between the spin-2 fluctuations $h'_{\mu\nu}$ and the scalar fluctuations as well, to resolve this we can apply a non-linear field redefinition 
\begin{equation}
	h^{I_k}_{\mu\nu} \rightarrow \phi^{I_k}_{\mu\nu} + \left(\nabla_{\mu} \nabla_{\nu} - \frac{1}{d+1} g_{\mu\nu} \square_{\text{AdS}}\right)\varphi^{I_k}\,,
\end{equation}
where
\begin{equation}
	 \varphi^{I_k} \mathcal X^{I_k} = \frac{\frac{1}{3d^2}(d+n-1)(dk + (d-1)(n-1))}{-\square_S + \frac{(n-1)^2(d-1)}{d^2} \epsilon^2 } s^{I_k} \mathcal X^{I_k}\,.
\end{equation}
At cubic order one finds that indeed $\phi_{\mu\nu}^{I_k}$ will be decoupled from $s^{I_k}$. Plugging back the scalar fluctuations $s_k$ into the original form \eqref{Eq: fluctuations} we find the following simple expressions
\begin{equation}
\begin{aligned}
	\delta^{(k)} g_{\mu\nu} &= \frac{d+n-1}{6(dk+n-1)\epsilon^2}  \left(\nabla_{\mu}\nabla_{\nu}- \frac{1}{d+1} g_{\mu\nu} \square_{\text{AdS}}\right) s^{I_k}  - k \frac{n-2}{3(d+1)} g_{\mu\nu} s^{I_k} \,,\nn\\
	\delta^{(k)} g_{\alpha\beta} &= \frac{k}{3} \, s^{I_k} \,.
\end{aligned}
\end{equation}
Out of these fluctuations we can extract the coefficients introduced in \eqref{eq: generic graviton fluctuation} to take the form
\begin{align}\label{Eq: coefs1}
  a_1(k)=\frac{2-n}{3(d+1)}k\,,\qquad a_2(k)=\frac{d+n-1}{6(dk+n-1) \epsilon^2}\,,\qquad a_3(k)= \frac{1}{3}k\,.
\end{align}
Finally, $a_4(k)$ is determined by starting with the fluctuation of $C_{n-1}$ and utilising the relation $G_7 = \star G_4 - \frac{1}{2}C_3\wedge G_4$, out of which we deduce that
\begin{align}
  \delta G_{d+1}\big|_{\text{AdS}_{d+1}} =   (-1)^d\frac{3k(\epsilon k - d)}{d} s_k \,\text{vol}_{\text{AdS}_{d+1}}\,,
\end{align}
and hence
\begin{align}\label{Eq: coefs2}
  a_4(k) = (-1)^d\frac{3k(\epsilon k - d)}{d}\,.
\end{align}
Specialising to the two main cases of interest we find that for AdS$_4\times S^7/\Z_{N_f}$
\begin{align}
  a_1(k) &= -\frac{5}{12}k\,,\quad a_2(k) = \frac{2}{k+2}\,,\quad a_3(k) = \frac{1}{3}k\,,\quad a_4(k) = -\frac{1}{2}k(k-6)\,,\nn\\
   b(k) &= \frac{1}{\kappa^2}\frac{k(k+3)(k-1)}{k+2}\,,
\label{eq: a and b AdS4}
\end{align}
while for AdS$_7\times S^4/\Z_2$ we have
 \begin{align}
  a_1(k) &= -\frac{2}{21}k\,,\quad a_2(k) = \frac{1}{8(2k+1)}\,,\quad a_3(k) = \frac{1}{3}k\,,\quad a_4(k) = k(k-3)\,,\nn\\
   b(k) &= \frac{1}{\kappa^2} \frac{k(k-1)(2k+3)}{4(2k+1)}\,.
  \label{eq: a and b AdS7}
\end{align}

\subsection{Geometric setup}

It is helpful to view $S^n$ as an embedded submanifold of $\R^{n+1}$, and use Cartesian coordinates $x_a$, $a=1,\dots,n+1$ on $\R^{n+1}$. Then, $S^n$ is defined by the constraint
\begin{align}
  x_1^2 + x_2^2 + x_3^2 +x_4^2 = 1-\rho^2,\qquad x_5^2 +\dots +x_{n+1}^2 = \rho^2\,,
\end{align}
where the coordinate $\rho$ goes over the range $\rho\in [0,1]$. In this way, we realise $S^n$ as a (singular) $S^3\times S^{n-4}$ fibration over an interval, where the $S^3$ degenerates at one end of the interval, and the $S^{n-4}$ at the other.

We are actually interested in an orbifold $S^n/\Z_k$ of $S^n$, where we identify points related by a rotation that stabilises the origin in the $\R^{n-3}$ spanned by $(x_5,\dots, x_{n+1})$. 
It is clear then that the fixed point locus of this orbifold action is the surface $\rho=0$, which is just a unit $S^3$.

The virtue of this formulation is that there is just one integral we will ever need. Namely, if $x_A$ are Cartesian coordinates in $\R^{D+1}$ and $S^D$ is defined by $x_A x_A = 1$, then
\begin{align}
  &\int_{S^D} d\Omega_D\,x_{A_1}x_{A_2}\dots x_{A_{2m}} \nn\\
  &\quad = \frac{\text{Vol}(S^D)}{(D+1)(D+3)\dots (D+2m-1)}\left(\delta_{A_1 A_2}\dots \delta_{A_{2m-1}A_{2m}} + \text{all other contractions}\right)		\nn\\
  &\quad = \frac{\pi^{(D+1)/2}}{2^{m-1}\Gamma\left(\frac{D+1}{2}+m\right)} \left(\delta_{A_1 A_2}\dots \delta_{A_{2m-1}A_{2m}} + \text{all other contractions}\right)\,.
\end{align}
Let us present some special cases of this formula that we'll use. Let $i=1,2,3,4$. Firstly on $S^n/\Z_k$ we have that for each $m=1,2,\dots$ and for any $f(\rho)$,
\begin{align}
  &\int_{S^n/\Z_k} d\Omega_p\, x_{i_1}x_{i_2}\dots x_{i_{2m}} f(\rho) \nn\\
  &\qquad = \frac{\pi^{(n+1)/2} \L_m[f]}{2^{m-2}\Gamma(\frac{n-3}{2})\Gamma(m+2)k}\left(\delta_{i_1i_2}\dots\delta_{i_{2m-1}i_{2m}} + \text{all other contractions}\right)  \,,
\label{eq: S5 gen int}
\end{align}
where
\begin{align}
  \L_m[f] = \int_0^{1} dt\, t^{n-4}(1-t^2)^{m+1}f(t)\,.
\end{align}
Secondly, now on the $S^3$ fixed point locus, we have
\begin{align}
  \int_{S^3}d\Omega_3\,  x_{i_1}x_{i_2}\dots x_{i_{2m}} = \frac{\pi^2}{2^{m-1}(m+1)!}\left(\delta_{i_1i_2}\dots\delta_{i_{2m-1}i_{2m}} + \text{all other contractions}\right)  \,.
\label{eq: S3 int1}
\end{align}
Finally, note that we ultimately want to describe our fields in terms of tensors of $SU(2)_L\times SU(2)_R$ rather than $SO(4)$. For this purpose, it is useful to define
\begin{align}
  x^{\bar{\alpha}\alpha} = (\sigma_i)^{\bar{\alpha}\alpha} x_i\,,
\end{align}
with $\bar{\alpha},\bar{\beta},\dots$ fundamental indices of $SU(2)_L$, and $\alpha,\beta,\dots $ fundamental indices of $SU(2)_R$. The $\sigma_i$ are the $SO(4)$ (i.e. 4-dimensional Euclidean) Pauli matrices, $(\sigma_i)^{\bar{\alpha}\alpha} = (i\sigma^P_1,i\sigma^P_2,i\sigma^P_3,\mathds{1}_2)$ in terms of the regular Pauli matrices $\sigma^P_a$. All we'll really need to know about these is that
\begin{align}
  (\sigma_i)^{\bar{\alpha} \alpha}(\sigma_i)^{\bar{\beta}\beta} = 2 \epsilon^{\bar{\alpha}\bar{\beta}} \epsilon^{\alpha\beta}\,.
\end{align}
We then straightforwardly rephrase the above integral identities as
\begin{align}
  &\int_{S^n/\Z_k} d\Omega_5\, x^{\bar{\alpha}_1 \alpha_1 }\dots x^{\bar{\alpha}_{2m} \alpha_{2m}} f(\rho^2) \nn\\
  &= \frac{4\pi^{(n+1)/2} \L_m[f]}{\Gamma(\frac{n-3}{2})\Gamma(m+2)\mathbf{\Delta}}\left(\epsilon^{\bar{\alpha}_1\bar{\alpha}_2} \epsilon^{\alpha_1\alpha_2}\dots \epsilon^{\bar{\alpha}_{2m-1}\bar{\alpha}_{2m}} \epsilon^{\alpha_{2m-1}\alpha_{2m}}  + \text{all other contractions}\right)  \,,
\label{eq: S5 gen int SU(2)}
\end{align}
on $S^n/\Z_k$, and 
\begin{align}
  \int_{S^3}d\Omega_3\,  x^{\bar{\alpha}_1 \alpha_1 }\dots x^{\bar{\alpha}_{2m} \alpha_{2m}} = \frac{2\pi^2}{(m+1)!}\left(\epsilon^{\bar{\alpha}_1\bar{\alpha}_2} \epsilon^{\alpha_1\alpha_2}\dots \epsilon^{\bar{\alpha}_{2m-1}\bar{\alpha}_{2m}} \epsilon^{\alpha_{2m-1}\alpha_{2m}}  + \text{all other contractions}\right)  \,,
\label{eq: S3 int}
\end{align}
on $S^3$.

\subsection{Decomposition of graviton modes on orbifold}\label{subsec: graviton modes}

A scalar spherical harmonic $s_k$ on $S^n$ of rank $k$ takes the form
\begin{align}
  s_k = s_{a_1a_2 \dots a_k} x_{a_1}x_{a_2}\dots x_{a_k}\,,
\end{align}
pulled back to $S^n$, where $s_{a_1\dots_k}$ is totally symmetric and traceless on any two indices, and recall $a=1,\dots,n+1$.

Acting with the orbifold breaks
\begin{align}
  SO(n+1)  \quad \longrightarrow \quad  SU(2)_L \times SU(2)_R \times H\,,
\end{align}
where $SO(4)\cong SU(2)_L\times SU(2)_R$ rotates the $x_i$, $i=1,2,3,4$ while $H$ acts on the $x_I$, $I=5,\dots,n+1$ and is the subgroup of $SO(n-3)$ preserved by the orbifold. This orbifold then projects out certain linear combinations of components of $s_{a_1a_2\dots a_k}$. From the remaining components we need to build irreducible representations of $SU(2)_L\times SU(2)_R\times H$.

Rather than doing this in full, note that the only graviton modes which can couple to gluon modes on the branes at the fixed point locus are precisely those which do not vanish at $\rho=0$. It is simple to see that all such modes are uncharged under $H$. So let's focus on these modes.

Then, we find straightforwardly that the scalar spherical harmonic corresponding to the operator $k_p$ is given by
\begin{align}
  k_p:\qquad s_{i_1 i_2 \dots i_{p-2}}x_{i_1}x_{i_2}\dots x_{i_{p-2}} (1-\rho^2)^{(k_p-p+2)/2} - (SO(n+1) \text{ traces})\,,
  \label{eq: kp mode}
\end{align}
where here $s_{i_1\dots i_{p-2}}$ is symmetric and traceless on its $SO(4)$ indices. Note that the rank $n$ symmetric traceless representation of $SO(4)$ is irreducible, and is nothing but the representation $(\frac{n}{2},\frac{n}{2})$ of $SU(2)_L\times SU(2)_R$. Thus, $k_p$ does indeed transform in the representation $(\frac{p}{2}-1,\frac{p}{2}-1)$ as required. As a consistency check, note that indeed the above formula makes sense only for
\begin{align}
  k_p-(p-2) = 0,2,4,\dots \,,
\end{align}
and that for $p=2,3$ the first of these options corresponds to $k_2=0$ and $k_3=1$, respectively, which as discussed above are pure gauge and so are thrown away. So we recover precisely the expected ranges for $p$ and $k_p$. 
 
To proceed, we will need to perform the removal of traces explicitly to get a closed form for the mode in (\ref{eq: kp mode}). There is a slick way to do this. It is easy to show that after removal of traces, the mode will take the form
\begin{align}
  k_p:\qquad s_{i_1 i_2 \dots i_{p-2}}x_{i_1}x_{i_2}\dots x_{i_{p-2}} (1-\rho^2)^{(k_p-p+2)/2} f_{k_p} \left(\frac{\rho^2 }{1-\rho^2}\right)\,,
  \label{eq: kp mode explicit1}
\end{align}
for a degree $(k_p-p+2)/2$ polynomial $f_{k_p}(x)$ we must determine. But the tracelessness condition is just the statement that the mode satisfies the Laplace equation $\nabla^2(\dots) = 0$ in the ambient space $\R^{n+1}$. This implies that $f_{k_p}(x)$ obeys
\begin{align}
  (k_p - p+2)(k_p+p)f_{k_p}(x) + 2\left(n-3-2k_px\right) f'_{k_p}(x) + 4 x (1+x)f''_{k_p}(x) = 0\,.
\end{align}
The solution to this equation we want is given by
\begin{align}
  f_{k_p}(x) &={}_2F_{1} \left(-\frac{k_p+p}{2}, \frac{p-k_p-2}{2},\frac{n-3}{2};-x\right)\,,
\end{align}
which is indeed a polynomial of degree $(k_p-p+2)/2$. Using some hypergeometric identities, we finally arrive at
\begin{align}
  &s_{(k_p)}(x,y)\nn\\
  &=\frac{1}{\N(k_p)}s_{\bar{\alpha}_1\dots \bar{\alpha}_{p-2},\alpha_1\dots \alpha_{p-2}}(y) x^{\bar{\alpha}_1\alpha_1}\dots x^{\bar{\alpha}_{p-2} \alpha_{p-2}}  \, {}_2 F_{1}\left(\frac{p+k_p+n-3}{2},\frac{p-k_p-2}{2},\frac{n-3}{2};\rho^2\right)\,,
  \label{eq: kp mode explicit}
\end{align}
where $y$ are coordinates in AdS$_{d+1}$. We have chosen to write everything in a manifestly $SU(2)_L\times SU(2)_R$ covariant way. So, $s_{\bar{\alpha}_1\dots \bar{\alpha}_{p-2},\alpha_1\dots \alpha_{p-2}}$ is totally symmetric on both its barred and unbarred indicies. $\N(k_p)$ is a normalisation factor that is in place to ensure that $s_{\bar{\alpha}_1\dots \bar{\alpha}_{p-2},\alpha_1\dots \alpha_{p-2}}(y)$ is a canonically normalised scalar field in AdS$_{d+1}$.

So now let's fix the normalisation factor. Using the integral (\ref{eq: S5 gen int SU(2)}), we compute
\begin{align}
  \N(k_p)^2 =  \frac{4\pi^{(n+1)/2}\Gamma\!\left(\frac{n-3}{2}\right) \Gamma\!\left(\frac{k_p-p+4}{2}\right)\Gamma\!\left(\frac{k_p+p+2}{2}\right)b(k_p) }{\epsilon^n k (p-1)(2k_p+n-1) \Gamma\!\left(\frac{k_p-p+n-1}{2}\right)\Gamma\!\left(\frac{k_p+p+n-3}{2}\right)}\,,
\label{eq: N}
\end{align}
where $b(k_p)$ is the function appearing in (\ref{eq: SUGRA kinetic term}) and determined for the cases of interest in (\ref{eq: a and b AdS4}) and (\ref{eq: a and b AdS7}). In arriving at this formula we've used the identity
\begin{align}
  &\int_0^1 dt\, t^{n-4}(1-t^2)^{p-1} \left[{}_2 F_{1}\left(\frac{p+k+n-3}{2},\frac{p-k-2}{2},\frac{n-3}{2};t^2\right)\right]^2 		\nn\\
  &\qquad = \frac{\Gamma\!\left(\frac{n-3}{2}\right)^2 \Gamma\!\left(\frac{k-p+4}{2}\right)\Gamma\!\left(\frac{k+p+2}{2}\right)}{(2k+n-1) \Gamma\!\left(\frac{k-p+n-1}{2}\right)\Gamma\!\left(\frac{k+p+n-3}{2}\right)}\,.
\end{align}
Finally, note that ${}_2F_{1}(\dots;0)$ and hence at the fixed point we have simply
\begin{align}
  s_{(k_p)}(x,y)&=\frac{1}{\N(k_p)}s_{\bar{\alpha}_1\dots \bar{\alpha}_{p-2},\alpha_1\dots \alpha_{p-2}}(y) x^{\bar{\alpha}_1\alpha_1}\dots x^{\bar{\alpha}_{p-2} \alpha_{p-2}}  \,.
\label{eq: sk mode}
\end{align}

\subsection{Decomposition of gluon modes}

We next consider the gluon modes. We once again adopt the strategy of working in a Cartesian embedding space, so as to utilise (\ref{eq: S3 int}).

These gluon modes are vector spherical harmonics on the $S^3$ fixed point locus at $\rho = 0$. We are interested in the modes transforming in the representation $\left(\tfrac{p}{2}-1,\tfrac{p}{2}\right)$ of $SU(2)_L\times SU(2)_R$ for $p=2,3,\dots$, which have scaling dimension $\Delta_p$.

The relevant vector spherical harmonics as 1-forms in $\R^4$ are then
\begin{align}
  V_p &=\frac{1}{\B(p)}\phi_{\bar{\alpha}_1\dots \bar{\alpha}_{p-2},\beta \gamma \alpha_1 \dots \alpha_{p-2}} x^{\bar{\alpha}_1 \alpha_1} \dots x^{\bar{\alpha}_{p-2} \alpha_{p-2}} \left((\sigma_j)^{\bar{\delta}(\beta}(\sigma_k)_{\bar{\delta}}{}^{\gamma)}\right)  x_j dx_k			\nn\\
  & =\frac{1}{\B(p)}\phi_{\bar{\alpha}_1\dots \bar{\alpha}_{p-2},\beta \gamma \alpha_1 \dots \alpha_{p-2}} x^{\bar{\alpha}_1 \alpha_1} \dots x^{\bar{\alpha}_{p-2} \alpha_{p-2}} (\sigma_j)^{\bar{\delta}\beta}(\sigma_k)_{\bar{\delta}}{}^{\gamma}   x_j dx_k	\,,
\label{eq: Vp mode}
\end{align}
where $\B(p)$ is a normalisation factor to ensure that the scalar field  $\phi_{\bar{\alpha}_1\dots \bar{\alpha}_{p-2}, \alpha_1 \dots \alpha_{p}} $ is canonically-normalised in AdS$_5$, which we will determine shortly.
Note that in the first line, the expression in brackets is nothing but the anti-self-dual 't Hooft matrix, written in a manifestly $SU(2)_R$ covariant form. In particular, one can check that
\begin{align}
  \left((\sigma_i)^{\bar{\delta}(\beta}(\sigma_j)_{\bar{\delta}}{}^{\gamma)}\right) = -\left((\sigma_j)^{\bar{\delta}(\beta}(\sigma_i)_{\bar{\delta}}{}^{\gamma)}\right) = -\frac{1}{2}\epsilon_{ijkl} \left((\sigma_k)^{\bar{\delta}(\beta}(\sigma_l)_{\bar{\delta}}{}^{\gamma)}\right) \,,
\end{align}
which will be vital when we come to compute the Wess-Zumino term. Note also that the asymmetry here means that automatically $V_p$ is tangent to $S^3$, i.e. $x_i (V_p)_i=0$.

So, to turn on the $p^\text{th}$ gluon mode, we simply set
\begin{align}
  A =  V_p^*\,,
\end{align}
where $V_p^*$ is the pullback of $V_p$ to $S^3$. Next, we need a few useful facts about $V_p^*$. Letting $a,b,\dots$ be abstract indices on $S^3$, we have
\begin{align}
  \nabla_b \nabla^b (V_p^*)_a 		&= (2-p^2) (V_p^*)_a	 \,,	\nn\\
  \epsilon_{abc} \nabla^b (V_p^*)^c 		&= -p (V_p^*)_a		\,,		\nn\\
  \nabla_a (V_p^*)^a						&= 0\,,
\end{align}
where the volume form appearing in the second expression is that of the unit round 3-sphere, and indeed all indices are raised and lowered using this unit round metric. The only other thing we need is the expression for the inner product of pulled-back 1-forms,
\begin{align}
  &g^{ab}(V_p^*)_a (V_q^*)_b \nn\\
  &\,\,= (V_p)_i (V_q)_i					\nn\\
  &\,\,= \frac{2}{\B(p)\B(q)} \phi^{(p)}_{\bar{\alpha}_1 \dots \bar{\alpha}_{p-2}, \gamma_1 \gamma_2 \alpha_1 \dots \alpha_{p-2}}\phi^{(q)}_{\bar{\beta}_1\dots \bar{\beta}_{q-2},}{}^{\gamma_1\gamma_2}{}_{\beta_1\dots \beta_{q-2}} x^{\bar{\alpha}_1\alpha_1}\dots x^{\bar{\alpha}_{p-2}\alpha_{p-2}} x^{\bar{\beta}_1\beta_1} \dots x^{\bar{\beta}_{q-2}\beta_{q-2}}	 \,.
\label{eq: Vp contraction}
\end{align}
So we're good to go. Let us first check the quadratic part of the gluon Lagrangian. We compute up to quadratic order,
\begin{align}
  &\text{Tr} \int_{S^3}\Big(-F\wedge \star_8 F\Big) \nn\\
  &\quad = \frac{1}{\B(p)^2}\frac{8\pi^2}{(p-1)}\text{Tr}\,\bigg( -\frac{1}{2\epsilon}\nabla_\mu \phi_{\bar{\alpha}_1 \dots \bar{\alpha}_{p-2},  \alpha_1 \dots \alpha_{p}} \nabla^\mu \phi^{\bar{\alpha}_1 \dots \bar{\alpha}_{p-2},  \alpha_1 \dots \alpha_{p}} \nn\\
  &\hspace{45mm}- \frac{\epsilon}{2}p^2 \phi_{\bar{\alpha}_1 \dots \bar{\alpha}_{p-2},  \alpha_1 \dots \alpha_{p}}  \phi^{\bar{\alpha}_1 \dots \bar{\alpha}_{p-2},  \alpha_1 \dots \alpha_{p}} +\dots\bigg) \,\text{vol}_{\text{AdS}_{d+1}}\,,
\end{align}
where we note the factors of $\epsilon$ coming from the  metric on $S^3$. We also then compute
\begin{align}
  \text{Tr} \int_{S^3}\Big(C_d\wedge F \wedge F\Big) = -\frac{d}{\B(p)^2}\frac{4\pi^2 p}{(p-1)}\text{Tr}\,\bigg( \phi_{\bar{\alpha}_1 \dots \bar{\alpha}_{p-2},  \alpha_1 \dots \alpha_{p}}  \phi^{\bar{\alpha}_1 \dots \bar{\alpha}_{p-2},  \alpha_1 \dots \alpha_{p}} +\dots\bigg)\,\text{vol}_{\text{AdS}_{d+1}}\,,
\end{align}
where we've used (\ref{eq: fluxes}).

We then have
\begin{align}
   &\frac{1}{g_\text{YM}^2} \,\text{Tr} \int_{S^3}\Big(-F\wedge \star_8 F + C_{d}\wedge F \wedge F\Big)\nn\\
   & \hspace{30mm}= \text{Tr}\,\bigg( -\frac{1}{2}\nabla_\mu \phi_{\bar{\alpha}_1 \dots \bar{\alpha}_{p-2},  \alpha_1 \dots \alpha_{p}} \nabla^\mu \phi^{\bar{\alpha}_1 \dots \bar{\alpha}_{p-2},  \alpha_1 \dots \alpha_{p}} \nn\\
   &\hspace{44mm}- \frac{1}{2}m_p^2 \phi_{\bar{\alpha}_1 \dots \bar{\alpha}_{p-2},  \alpha_1 \dots \alpha_{p}}  \phi^{\bar{\alpha}_1 \dots \bar{\alpha}_{p-2},  \alpha_1 \dots \alpha_{p}} +\dots\bigg)\,\text{vol}_{\text{AdS}_{d+1}}\,,
\end{align} 
where
\begin{align}
  m_p^2 = \Delta_p(\Delta_p-d),\qquad \Delta_p = \left(\frac{d-2}{2}\right)p\,,
\end{align}
as expected, and we have fixed the normalisation
\begin{align}
  \B(p)^2 = \frac{1}{g_\text{YM}^2}\frac{8\pi^2 }{\epsilon(p-1)} \,.
\label{eq: B norm}
\end{align}

\subsection{Computation of cubic coupling}

Let's finally compute the cubic coupling. For certainty, the AdS$_{d+1}$ action of the modes we're interested in up to cubic order is
\begin{align}
  S = S_\text{kin} + S_\text{cubic}\,.
\end{align}
The kinetic term is
\begin{align}
  S_\text{kin} &= \int_{\text{AdS}_{d+1}} d^5 x \sqrt{-g_{\text{AdS}}} \,\,\Bigg(  \sum_{p=2}^\infty \sum_{k_p} \bigg( -\frac{1}{2} \nabla_\mu(s_{(k_p)})_{\bar{\alpha}_1\dots \bar{\alpha}_{p-2},\alpha_1 \dots \alpha_{p-2}}\nabla^\mu (s_{(k_p)})^{\bar{\alpha}_1\dots \bar{\alpha}_{p-2},\alpha_1 \dots \alpha_{p-2}}		\nn\\
  &\hspace{60mm}-\frac{1}{2}\Delta_{k_p}(\Delta_{k_p}-d)(s_{(k_p)})_{\bar{\alpha}_1\dots \bar{\alpha}_{p-2},\alpha_1 \dots \alpha_{p-2}} (s_{(k_p)})^{\bar{\alpha}_1\dots \bar{\alpha}_{p-2},\alpha_1 \dots \alpha_{p-2}} \bigg)		\nn\\
  &\hspace{42mm} + \text{Tr} \sum_{p=2,3,\dots } \bigg( -\frac{1}{2} \nabla_\mu(\phi_{(p)})_{\bar{\alpha}_1\dots \bar{\alpha}_{p-2},\alpha_1 \dots \alpha_{p}}\nabla^\mu (\phi_{(p)})^{\bar{\alpha}_1\dots \bar{\alpha}_{p-2},\alpha_1 \dots \alpha_{p}}	\nn\\
  &\hspace{65mm} -\frac{1}{2}\Delta_p(\Delta_p-d) (\phi_{(p)})_{\bar{\alpha}_1\dots \bar{\alpha}_{p-2},\alpha_1 \dots \alpha_{p}} (\phi_{(p)})^{\bar{\alpha}_1\dots \bar{\alpha}_{p-2},\alpha_1 \dots \alpha_{p}} \bigg) \Bigg)\,,
\end{align}
where in the first line the sum over $k_p$ runs over the values
\begin{align}
  k_p = \left\{ \begin{aligned}
  	\,\, p, p+2, p+4,\dots		\qquad & p=2,3			\\
  	\,\, p-2,p,p+2,\dots 		\qquad & p>3
  \end{aligned} \right. \,,
\label{eq: kp range}
\end{align}
and the trace is over the flavor group.

Meanwhile, the cubic term is
\begin{align}
  S_\text{cubic}  &= \text{Tr}\int_{\text{AdS}_{d+1}} d^{d+1} x \sqrt{-g_{\text{AdS}}} \,\,\nn\\
  &\quad \times \Bigg(\sum_{p, q=2}^\infty \sum_r \sum_{k_r} \frac{1}{2} \beta_{pqk_r} (\phi_{(p)})_{\bar{\alpha}_1\dots \bar{\alpha}_{(p+q-r-2)/2}\bar{\beta}_{1}\dots \bar{\beta}_{(p-q+r-2)/2},\gamma \delta \alpha_1\dots \alpha_{(p+q-r-2)/2}\beta_{1}\dots \beta_{(p-q+r-2)/2}} 		\nn\\
  &\hspace{48mm}\times (\phi_{(q)})^{\bar{\alpha}_1\dots \bar{\alpha}_{(p+q-r-2)/2}}{}_{\bar{\beta}_{(p-q+r)/2}\dots \bar{\beta}_{r-2},}{}^{\gamma \delta \alpha_1\dots \alpha_{(p+q-r-2)/2}}{}_{\beta_{(p-q+r)/2}\dots \beta_{r-2}} 			\nn\\
  &\hspace{48mm}\times (s_{(k_r)})^{\bar{\beta}_1 \dots \bar{\beta}_{r-2},\beta_1\dots \beta_{r-2}} \Bigg) \,,
\label{eq: S cubic}
\end{align}
where the sum over $r$ runs over the values
\begin{align}
  r = |p-q|+2,|p-q|+4,\dots,p+q-2\,,
\label{eq: r values}
\end{align}
and the sum over $k_r$ runs over the values (\ref{eq: kp range}). Then $\beta_{pqk_r}$ receives contributions from both the Yang-Mills and Wess-Zumino terms. These are
\begin{align}
  &\left[\text{Tr} \int\Big(-F\wedge \star_{d+4} F\Big)\right]_{pqk_r} \nn\\
  &\qquad = \text{Tr}\int d^{d+4}x\sqrt{-g} \,\bigg( \epsilon k_r pq(a_3-(d+1)a_1) \,s_{(k_r)} (V^*_p)^a (V_q^*)_a		\nn\\
  &\hspace{80mm}+ \epsilon^{-1}\left((1-d)a_1-a_3\right)\,s_{(k_r)} \nabla_\mu(V^*_p)^a \nabla^\mu (V_q^*)_a		\nn\\
  &\hspace{80mm}+ 4\epsilon^{-1} a_2 \, \nabla_\mu \nabla_\nu s_{(k_r)} \nabla^\mu(V^*_p)^a \nabla^\nu (V_q^*)_a		\nn\\
  &\hspace{80mm} - \frac{4a_2}{d+1}\epsilon^{-1} \nabla_\mu \nabla^\mu s_{(k_r)} \nabla^\mu(V^*_p)^a \nabla_\mu (V_q^*)_a \bigg)\,,
\end{align}
and
\begin{align}
  &\left[\text{Tr} \int\Big(C_d\wedge F \wedge F\Big)\right]_{pqk_r} \nn\\
  &\qquad = \text{Tr}\int d^{d+4}x\sqrt{-g} \,\bigg( 2(-1)^{d+1}a_4 \Big(q(V_q^*)^a \nabla^\mu s_{(k_r)}\nabla_\mu (V_p^*)_a+ p(V_p^*)^a \nabla^\mu s_{(k_r)}\nabla_\mu (V_q^*)_a\Big)\bigg)\,,
\end{align}
where in each expression the volume form is that of the pullback to AdS$_{d+1}\times S^3$ of the metric (\ref{eq: metric}). To evaluate these term, we need a few identities. For scalar functions $f_1,f_2,f_3$, we have 
 \begin{align}
  \nabla_\mu \nabla_\nu f_1 \nabla^\mu f_2 \nabla^\nu f_3 &= \frac{1}{4}\Big((\square^2 f_1) f_2 f_3 -  f_1 (\square^2 f_2)f_3 - f_1 f_2 (\square^2 f_3) + 2 f_1 (\square f_2)(\square f_3)\Big) + \nabla_\mu \left(\dots \right)\,,		\nn\\
   f_1 \nabla^\mu f_2 \nabla_\mu f_3 &= \frac{1}{2}\Big((\square f_1) f_2 f_3 -  f_1 (\square f_2)f_3 - f_1 f_2 (\square f_3)\Big) + \nabla_\mu \left(\dots \right)	\,.
\end{align}
The resulting expressions appear complicated, but undergo a remarkable simplification when we plug in in the values (\ref{eq: a and b AdS4}) or (\ref{eq: a and b AdS7}) for the relevant cases. In the end we find 
\begin{align}
 &\left[\text{Tr} \int\Big(-F\wedge \star_8 F + C_{d}\wedge F \wedge F\Big)\right]_{pqk_r} \nn\\
 &\quad =  \left(\frac{2^{5-n}(k_r+p-q)(k_r+q-p)(k_r+p+q-2)(k_r+p+q-n+1)}{(d-2)k_r+2}\right) 	\nn\\
 &\hspace{15mm} \times \text{Tr}\int d^{d+4}x\sqrt{-g} \,\Big(s_{(k_r)} (V^*_p)^a (V_q^*)_a\Big)\,.
\end{align}
We finally substitute in the explicit forms of the graviton harmonics in (\ref{eq: sk mode}) along with the expression (\ref{eq: Vp contraction}) for the contraction of two gluon modes, and perform the integral over $S^3$ using (\ref{eq: S3 int}). The end result is the cubic coupling
\begin{align}
  \beta_{pqk_r} &= \frac{1}{g_\text{YM}^2 \N(k_r)\B(p)\B(q)} \left(\frac{2^{5-n}(k_r+p-q)(k_r+q-p)(k_r+p+q-2)(k_r+p+q-n+1)}{(d-2)k_r+2}\right)		\nn\\
  &\qquad \times \left(\frac{4\pi^2\Gamma(p-1)\Gamma(q-1)\Gamma(r-1)}{\, \Gamma\!\left(\frac{p+q-r}{2}\right)\Gamma\!\left(\frac{r+p-q}{2}\right)\Gamma\!\left(\frac{r+q-p}{2}\right)\Gamma\!\left(\frac{r+p+q-2}{2}\right) }\right)\,,
\end{align}
where the expression in the final line comes from the integral over $S^3$. Plugging in the normalisation coefficients $\N(k_r)$ from (\ref{eq: N}) and $\B(p)$ from (\ref{eq: B norm}) (which in particular kills the dependence on $g_{\text{YM}}^2$), we land on the final answer (\ref{Eq: cubic coupling general d n}).

\section{Exchange diagrams in Mellin space}
\label{blockDets}
In this appendix, we give the full details of the exchange Witten diagrams in Mellin space, and how to write them as reduced blocks. We include many of the central results of this appendix in the attached \texttt{Mathematica} notebook. In particular, the operators $\mathcal{D}_J^\epsilon$, and $\widetilde{\mathcal{D}}_J^\epsilon$, as well as our expressions for the Mack polynomials, can be found in this notebook.

\subsection{Superconformal Ward identities in Mellin space}
The superconformal Ward identities encode the action of the fermionic generators of the superconformal group. In our conventions, they take a universal form in position space~\cite{Dolan:2004mu,Baume:2019aid}:
\begin{equation}
    (z\partial_z-\epsilon \alpha\partial_\alpha)G(z,\bar z;\alpha)\Bigg|_{\alpha=\frac{1}{z}}=0\,,
\end{equation}
along with a second equation given by inverting $z\leftrightarrow\bar{z}$ above. In even dimensions, this equation has a simple solution, of the form:
\begin{equation}
    G(z,\bar z;\alpha) = f_{\text{disc}}(z,\bar z;\alpha) + R \circ H(z,\bar z;\alpha)\,,\label{eq:position_scwi_solution}
\end{equation}
where $f_{\text{disc}}$ is the disconnected contribution, $R$ is a local differential operator and $H$ is known as a reduced amplitude. However, in odd dimensions, the operator $R$ is non-local, and this approach becomes difficult to work with.

After Mellin transforming the amplitude, however, the Ward identity always admits a decomposition of this form, regardless of the dimension~\cite{Virally:2025nnl}, and we are able to define a reduced Mellin amplitude even in $d=3,5$.

In order to obtain the Ward identities in Mellin space, we follow~\cite{Zhou:2017zaw}. First, we use the chain rule to write:
\begin{equation}
    z\partial_z = U\partial_U + V\partial_V - \frac{1}{1-z}V\partial_V\,,
\end{equation}
but do not evaluate the derivatives. We can then act with the $\alpha\partial_\alpha$ derivative, and take $\alpha\to\frac{1}{z}$. Multiplying by $z^a(1-z)$, where $a$ is the degree of $G(z,\bar z;\alpha)$ as a polynomial in $\alpha$ and adding the equation obtained by taking $z\leftrightarrow \bar{z}$, we can always rewrite the resulting expression in terms of integer powers of $U,V,U\partial_U,V\partial_V$ acting on $G(z,\bar z;\alpha)$. Then, we can use the following dictionary to rewrite these as operators in Mellin space, which follows from our definition of the Mellin transform~\eqref{mellin}:
\begin{align}
    \begin{split}
        &U\partial_U\to \frac{s}{2}\,,\quad V\partial_V\to \frac{t-\Delta_2-\Delta_3}{2}\,,\\
        U^mV^n\to&\left(\frac{\Delta_1+\Delta_2-s}{2}\right)_m\left(\frac{\Delta_3+\Delta_4-s}{2}\right)_m\left(\frac{\Delta_1+\Delta_4-t}{2}\right)_n\left(\frac{\Delta_2+\Delta_3-t}{2}\right)_n\\
        &\times\left(\frac{\Delta_1+\Delta_3-u}{2}\right)_{-m-n}\left(\frac{\Delta_2+\Delta_4-u}{2}\right)_{-m-n}P_{-2m,-2n}\,,
    \end{split}
\end{align}
where $\Delta_i$ are the scaling dimensions of the four external operators, and we defined the shift operator $P_{m,n}$, which acts on functions of $s$ and $t$ as:
\begin{equation}
    P_{m,n}f(s,t) = f(s+m,t+n)\,.
\end{equation}
We now turn to the question of solving the Ward identities. In both $\langle 22pp\rangle$ and $\langle 2p2p\rangle$ correlators, we will see that the solutions are quadratic in $\alpha$\footnote{More precisely, in $\langle 2p2p\rangle$ correlators, the solution is $\alpha^{\frac{p}{2}-1}$ multiplying something quadratic in $\alpha$.}. In even dimensions, the position space operator $R$ in~(\ref{eq:position_scwi_solution}) is also quadratic in $\alpha$, which means that the reduced amplitudes have no dependence on the R-symmetry cross-ratio.

In order to solve the superconformal Ward identities in Mellin space, we assume that there exists a similar object in Mellin space, an operator $\mathcal{D}_J^\varepsilon$ which acts on any function of $s,t$ and gives a solution to the Ward identities. In order to find this operator, we will follow the procedure of~\cite{Virally:2025nnl}.

When translated into Mellin space, the position operators $R$ in even dimensions are of the form of shift operators with polynomial prefactors, so we assume that the answer is of this form in all dimensions. We then note that shift operators are diagonalized by power laws, i.e.:
\begin{equation}
    P_{m,n} X^sY^t = (X^mY^n)X^sY^t\,.
\end{equation}
Therefore, we start by assuming that there exists a function which solves the Ward identities of the form:
\begin{equation}
    F_{\mathcal{D}}(s,t;\alpha) = \sum_{a,b,m,n,J} \lambda_{m,n,a,b,J}X^mY^n s^at^b\alpha^J X^sY^t\,.
\end{equation}
Since this function is a polynomial in all relevant variables, multiplying the overall powers $X^sY^t$, it is generally simple to find such solutions.

We choose the lowest-order solution to the Ward identities of the form above. Then, if there exists an operator $\mathcal{D}$ with the desired properties, it will be the case that:
\begin{equation}
    F_{\mathcal{D}}(s,t;\alpha) = \sum_J\mathcal{D}_J^\varepsilon\alpha^J X^sY^t\,.
\end{equation}
Since shift operators are diagonalized when acting on $X^sY^t$, it is easy to read off the form of $\mathcal{D}$ from this single solution to the Ward identities:
\begin{equation}
    \mathcal{D}_J^\epsilon =  \sum_{a,b,m,n}\lambda_{a,b,m,n,J} s^at^b P_{m-N,n-N}\,,
\end{equation}
where $N$ is the largest combined power of $X$ and $Y$ in the solution above. We can then show, using this definition, that
\begin{equation}
    F(s,t;\alpha) = \sum_J \mathcal{D}_J^\epsilon\alpha^J f(s,t)
\end{equation}
is always a solution to the Ward identities, for any function $f(s,t)$, which proves that such an operator $\mathcal{D}$ exists and takes this form.

This operator is not unique. In fact, it is only defined up to what is called a ``Mellin ambiguity'' in~\cite{Virally:2025nnl}, which is the right-multiplication of $\mathcal{D}_J^\varepsilon$ by any operator independent of $J$. Here, we choose to take the simplest choice of $\mathcal{D}_J^\varepsilon$ with polynomial coefficients. The subtraction of $N$ from the shift operators is a choice of a Mellin ambiguity, an overall shift, which makes all shift operators act in the negative direction, mirroring the solutions from position space. With this choice of the ambiguity, we find the following operators for the $\langle 22pp\rangle$ correlator:
\begin{equation}
    \begin{aligned}
        \mathcal{D}_0^\epsilon ={}& (-(p+2) \epsilon +s+t+2) (-(p+2)\epsilon +s+t)^2P_{0,0}\\ 
        &+(2 p\epsilon -s) (-(p+4) \epsilon +s+t+2)(-(p+2) \epsilon +s+t)P_{-2,0}\\
        &+((p+2) \epsilon -t) (-(p+4) \epsilon+s+t+2) (-(p+2) \epsilon +s+t)P_{0,-2}\,,\\
        \mathcal{D}_1^\epsilon={}&-(-(p+2) \epsilon +s+t)^2 (-(p+2)\epsilon +s+t+2)P_{0,0}\\
        &+(t-(p+2) \epsilon )(-2 (p+2) \epsilon +s+2 t) (-(p+2)\epsilon +s+t)P_{0,-2}\\
        &+(s-2 p\epsilon ) (-(p+2) \epsilon +t+4)(-(p+2) \epsilon +s+t)P_{-2,0}\\
        &+(s-2 p \epsilon ) (t-(p+2)\epsilon ) ((p-2) \epsilon+s-t)P_{-2,-2}\\
        &+(s-4\epsilon ) (-2 p \epsilon +s-2) (s-2p \epsilon )P_{-4,0}\\
        &+((p+2) \epsilon-t+2) (t-(p+2) \epsilon )^2P_{0,-4}\,,\\
        \mathcal{D}_2^\varepsilon={}&(s-2 (\epsilon +1)) (s-2 p\epsilon ) ((p+2) \epsilon-t)P_{-2,-2}\\
        &+ (s-2 (\epsilon +1)) (s-2p \epsilon ) (-(p+2) \epsilon+s+t)P_{-2,0}\\
        &-\left( (s-4 \epsilon )(-2 p \epsilon +s-2) (s-2 p \epsilon)\right)P_{-4,0}\,.
    \end{aligned}
\end{equation}
For the $\langle 2p2p\rangle$ correlator, we have:
\begin{equation}
    \begin{aligned}
        \widetilde{\mathcal{D}}_{\frac{p}{2}-1}^\epsilon={}&(s+t-4 \epsilon )(s+t-4 \epsilon +2) (-2 p \epsilon+s+t)P_{0,0}\\ &((p+2) \epsilon -s) (s+t-4\epsilon ) (-2 (p+1) \epsilon+s+t+2)P_{-2,0}\\
        &+ ((p+2) \epsilon -t)(s+t-4 \epsilon ) (-2 (p+1) \epsilon+s+t+2)P_{0,-2}\\
        \widetilde{\mathcal{D}}_{\frac{p}{2}}^\epsilon={}&-(s+t-4\epsilon ) (s+t-4 \epsilon +2) (-2 p\epsilon +s+t)P_{0,0}\\
        &+ ((p+2)\epsilon -s) ((p+2) \epsilon -t-4)(s+t-4 \epsilon )P_{-2,0}\\
        &+ (t-(p+2)\epsilon ) (s+t-4 \epsilon ) (-(3p+2) \epsilon +s+2 t)P_{0,-2}\\
        & +(s-t) ((p+2) \epsilon -s)((p+2) \epsilon -t)P_{-2,-2}\\
        &+(-(p+2)\epsilon +s-2) (s-(p+2) \epsilon)^2P_{-4,0}\\
        &+(t-(p+2) \epsilon )^2((p+2) \epsilon -t+2)P_{0,-4}\\
        \widetilde{\mathcal{D}}_{\frac{p}{2}+1}^\epsilon={}&(p \epsilon -s+2) ((p+2)\epsilon -s) (s+t-4 \epsilon)P_{-2,0}\\
        &+(p \epsilon -s+2) ((p+2)\epsilon -s) ((p+2) \epsilon-t)P_{-2,-2}
        \\
        &+((p+2) \epsilon -s+2)(s-(p+2) \epsilon )^2P_{-4,0}\,.
    \end{aligned}
\end{equation}
In the attached \texttt{Mathematica} notebook, these operators can be found both in the basis of powers of $\alpha$, as in the last two equations, and in the basis of the $\mathcal{Y}^p_J(\alpha)$, as in \eqref{redM}.

\subsection{Mack polynomials}
We conclude this appendix by giving our conventions for the Mack polynomials, from which we build the bosonic exchange diagrams $M^{\Delta_{12},\Delta_{34}}_{\Delta,\ell}(s,t)$:
\begin{equation}
    M^{\Delta_{12},\Delta_{34}}_{\Delta,\ell}(s,t)=\sum_{m=0}^\infty\frac{K_{\Delta,\ell,m}^{\Delta_1,\Delta_2,\Delta_3,\Delta_4}Q^{\Delta_{12},\Delta_{34}}_{\Delta,\ell,m}(u-\Delta_1-\Delta_4)}{s-(\Delta-\ell+2m)}\,,
\end{equation}
where
\begin{equation}
    K^{\Delta_1,\Delta_2,\Delta_3,\Delta_4}_{\Delta,\ell,m} = -\frac{2^{1-\ell } (\ell +\tau -1)_{\ell } \Gamma (2
   \ell +\tau )}{\Gamma _{12}^2 \Gamma _{34}^2 m!
   \Gamma \left(-m-\frac{\tau }{2}+\frac{\Delta
   _1}{2}+\frac{\Delta _2}{2}\right) \Gamma
   \left(-m-\frac{\tau }{2}+\frac{\Delta
   _3}{2}+\frac{\Delta _4}{2}\right)
   \left(-\frac{d}{2}+\ell +\tau +1\right)_m}
\end{equation}
and
\begin{equation}
    \Gamma_{ab}^2 = \Gamma\left(\ell+\frac{\tau+\Delta_a-\Delta_b}{2}\right)\Gamma\left(\ell+\frac{\tau-\Delta_a+\Delta_b}{2}\right)\,,
\end{equation}
and $\tau=\Delta-\ell$ is the twist. For the polynomials $Q^{a,b}_{\Delta,\ell,m}(u)$, we found the following representation, which is a generalization of the result of~\cite{Dey:2017fab}, useful:
\begin{equation}
    \begin{aligned}
        &Q_{\Delta,\ell,m}^{a,b}(u) = \sum_{k=0}^\ell\sum_{n=0}^{\ell-k}(-m)_k\left(m+\frac{u+\tau}{2}\right)_n\mu(\ell,k,n,\tau,d,a,b)\,,\\
        &\mu(\ell,k,n,\tau,d,a,b)=\frac{2^{\ell } \ell ! (-1)^{k+n+\ell } \Gamma (\ell+\tau -1) (\ell +\tau -1)_n}{k! n! \Gamma (2 \ell+\tau -1) (-k-n+\ell )!}\\
        &\times \left(\frac{1}{2}(-a+b+d+2 n-2)\right)_k\left(-\frac{a}{2}+k+n+\frac{\tau}{2}\right)_{-k-n+\ell }\left(\frac{b}{2}+k+n+\frac{\tau}{2}\right)_{-k-n+\ell}\\
        &\times\, _4F_3\left(-k,3-d-n-\ell,\frac{2+a-d+\tau}{2},\frac{2-b-d+\tau}{2};\frac{a-b-d}{2}-k-n+2,2-\frac{d}{2}-\ell,2-d+\tau;1\right)\,.
    \end{aligned}
\end{equation}

\bibliographystyle{JHEP}
\bibliography{gravitonMtheory}

\end{document}